\documentclass[a4paper,11pt]{article}
\pdfoutput=1 

\usepackage{jheppub} 

\usepackage[T1]{fontenc} 

\usepackage{epsfig}
\usepackage{amsfonts}
\usepackage{latexsym}
\usepackage{amsmath}
\usepackage{amssymb}
\usepackage{slashed}
\usepackage{longtable}
\usepackage{comment}
\usepackage{subfigure}

\usepackage{graphicx}

 \def\p{\partial}

\newcommand{\bea}{\begin{eqnarray}}
\newcommand{\eea}{\end{eqnarray}}
\newcommand{\be}{\begin{equation}}
\newcommand{\ee}{\end{equation}}
\newcommand{\ba}{\begin{align}}
\newcommand{\ea}{\end{align}}

\def\Or[#1]{{\text{O}}\left({#1}\right)}
\def\dotl[#1,#2]{\left\langle #1, #2 \right\rangle}
\def\dotlb[#1,#2]{[ #1, #2 ]}
\def\dotp[#1,#2]{(#1) \cdot (#2)}
\def\aff[#1,#2]{\hat{#1}(#2)}
\def\n4sym{{\cal N}=4 SYM}
\def\>{\rangle}
\def\<{\langle}
\def\weight[#1,#2,#3]{\{(#1),#2,#3\}}
\def\ads[#1]{$\text{AdS}_{#1}$}

\title{\boldmath{Magnetotransport in multi-Weyl semimetals: A kinetic theory approach }}

\author{Renato M. A. Dantas,}
\author{Francisco Pe\~{n}a-Benitez,}
\author{Bitan Roy,}
\author{Piotr Sur\'owka}

\affiliation{Max-Planck-Institut  f\"ur Physik komplexer Systeme, N\"othnitzer Str. 38, 01187 Dresden, Germany}

\emailAdd{rmad@pks.mpg.de}
\emailAdd{pena@pks.mpg.de}
\emailAdd{bitanroy@pks.mpg.de}
\emailAdd{surowka@pks.mpg.de}

\abstract{
We study the longitudinal magnetotransport in three-dimensional multi-Weyl semimetals, constituted by a pair of (anti)-monopole of arbitrary integer charge ($n$), with $n=1,2$ and $3$ in a crystalline environment. For any $n>1$, even though the distribution of the underlying Berry curvature is \emph{anisotropic}, the corresponding intrinsic component of the longitudinal magnetoconductivity (LMC), bearing the signature of the chiral anomaly, is \emph{insensitive} to the direction of the external magnetic field ($B$) and increases as $B^2$, at least when it is sufficiently weak (the semi-classical regime). In addition, the LMC scales as $n^3$ with the monopole charge. 
We demonstrate these outcomes for two distinct scenarios, namely when inter-particle collisions in the Weyl medium are effectively described by (a) a single and (b) two (corresponding to inter- and intra-valley) scattering times. While in the former situation the contribution to LMC from chiral anomaly is inseparable from the non-anomalous ones, these two contributions are characterized by different time scales in the later construction. Specifically for sufficiently large inter-valley scattering time the LMC is dominated by the anomalous contribution, arising from the chiral anomaly. 
The predicted scaling of LMC and the signature of chiral anomaly can be observed in recently proposed candidate materials, accommodating multi-Weyl semimetals in various solid state compounds.      
}


\arxivnumber{1802.07733}

\begin{document} 
\maketitle
\flushbottom

\section{Introduction}

Quantum phenomena can have macroscopic manifestations, such as the anomaly-induced transport in systems, constituted by linearly dispersing massless Weyl fermions in three dimensions. The most celebrated ones are the chiral magnetic and chiral vortical effects~\cite{PhysRevD.20.1807,PhysRevD.22.3080,Kharzeev,Erdmenger:2009ky,Banerjee:2011kta}, both being intimately related with quantum anomalies~\cite{Son:2009tf,Landsteiner:2011cp}. The non-dissipative current describing these phenomena is given by 
\begin{equation}
\mathbf J = \sigma_{_B} \; {\mathbf B} + \sigma_{\omega} \; {\boldsymbol \omega},
\end{equation}
where $\mathbf B$ and $\boldsymbol{\omega}$ are magnetic and vorticity fields, respectively. Here, $\sigma_{_B}$ and $\sigma_\omega$ correspond to the chiral magnetic and chiral vortical conductivity, respectively.

\begin{figure}[t!]
\centering
 \includegraphics[width=0.8\textwidth]{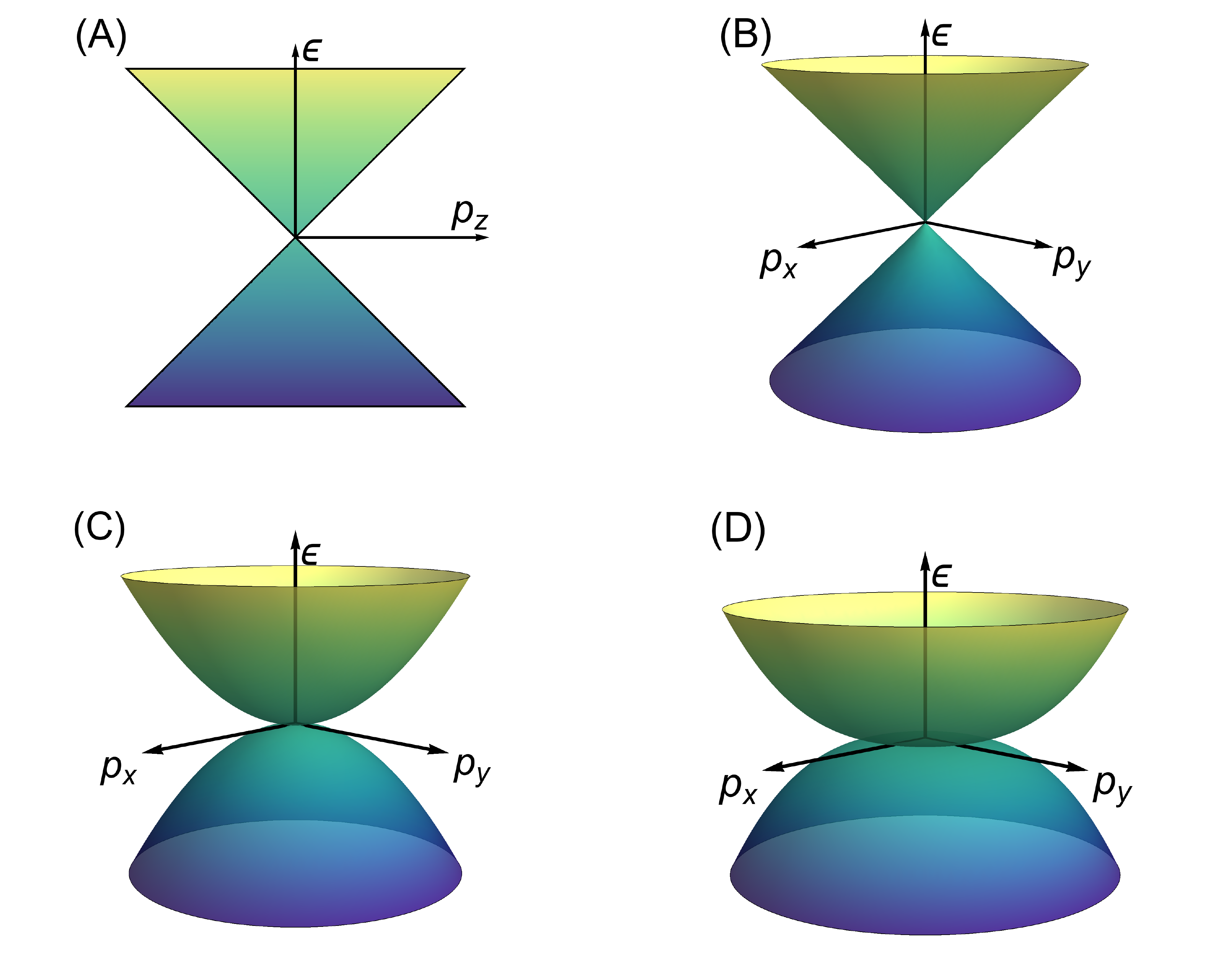}
\caption{Quasiparticle spectra in a multi-Weyl system along various high-symmetry directions in the close vicinity of Weyl nodes, characterized by an integer monopole charge $n$. Note that dispersion always scales linearly with $p_z$ (see panel A) for any $n$. But, in the $p_x-p_y$ plane the energy scales as $E \sim |p_\perp|$ for $n=1$ (see panel B), $E \sim |p_\perp|^2$ for $n=2$ (see panel C) and $E \sim |p_\perp|^3$ for $n=3$ (see panel D), where $p_\perp=\sqrt{p^2_x+p^2_y}$. A rotational symmetry is always present in the $p_x-p_y$ plane. Here, momentum ${\bf p}=\left( p_x, p_y, p_z\right)$ is measured from the Weyl node. Note that for $n>1$ the system looses the Lorentz invariance.   
}~\label{spectra}
\end{figure}

At the classical level, massless left- and right-handed Weyl spinors separately exhibit chiral symmetries and independent rotations of the phase can be performed for each species. By contrast, at the quantum level at most one of these rotations can be preserved (leaving the path-integral action invariant), a phenomenon known as \emph{quantum anomaly}. In particular, the electromagnetic gauge invariance requires $U(1)_e=U(1)_L+U(1)_R$ to be preserved, while its chiral counterpart $U(1)_5=U(1)_L-U(1)_R$ suffers an anomalous violation~\cite{Bertlmann:1996xk, fujikawa}. In three spatial dimensions  there may be two different sources to the anomalous non-conservation of the chiral charge. The first one, the so-called pure gauge anomaly, is present when parallel electric and magnetic field are switched on in the system \cite{Adler, Bell:1969ts}. The second one is called mixed gauge-gravitational anomaly and violates the conservation of chiral charge when the system is placed on a curved background~\cite{DELBOURGO1972381}. 
Even though signatures of these anomalies can be found in transport, for concreteness we only consider the imprint of pure gauge anomaly in multi-Weyl systems~\footnote{For a detailed description of the effect of the mixed-gauge gravitational anomaly on transport coefficients see Refs.~\cite{Landsteiner:2011cp,Lucas:2016omy,Landsteiner:2016stv}. For an experimental signature of such an anomaly in a Weyl semimetal consult Ref.~\cite{Gooth:2017mbd}.}.
In the high-energy physics such an effect is expected to be present in the quark-gluon plasma, experimentally created in heavy-ion collisions  (see Ref.~\cite{Liao:2016diz} and references therein). Furthermore, chiral anomaly leaves its signature in condensed matter systems, accommodating emergent Weyl quasiparticles at low-energies~\cite{ArmitageReview}.
Due to a \emph{no-go theorem}~\cite{nielsen}, Weyl fermions are always realized in pairs (except on the surface of a time-reversal invariant four-dimensional topological insulators) and each copy can be classified according to its \emph{chirality}: left or right. On the other hand, it has been shown that chiral magnetic effect vanishes in  Weyl semimetals at equilibrium (for a detailed discussion see \cite{Landsteiner2013,Kharzeev2013,Vazifeh2013}). Therefore, nonequilibrium signatures need to be explored in order to measure the effects of anomalies in Weyl materials.

Any \emph{non-orthogonal} arrangement of the electric ${\bf E}$ and magnetic ${\bf B}$ fields (such that ${\bf E} \cdot {\bf B} \neq 0$) causes a violation of the conservation of chiral charge. But, for the sake of concreteness, we restrict ourselves to the situation where the external electric and magnetic fields are always parallel to each other. We show that the system then becomes more conductive with an increasing magnetic field, an effect often refered as \emph{negative longitudinal magnetoresistance} (LMR), a hallmark signature of the Adler-Jackiw-Bell chiral anomaly~\cite{NielsenABJ}. Such an observation should be contrasted with the situation in a normal metal, without any Berry curvature, where magnetoresistence is typically \emph{positive}.

 In the language of condensed matter physics, the Weyl nodes, where Kramers non-degenerate valence and conduction bands touch each other, act as source and sink of Abelian Berry flux or curvature. Typically,  Weyl points with different chiralities are separated in the momentum space. Otherwise, such defects in the reciprocal space can be characterized with an integer monopole number that in turn also defines the topological invariant of the system (see Appendix~\ref{ap:topological}). In fact the Berry flux and quantum anomalies are directly connected \cite{Son:2012wh}, as we demonstrate here for multi-Weyl semimetals (see Ref.~\cite{PhysRevB.96.085112}).

So far, both theoretical~~\cite{Son&Spivak, Onoda, Kharzeev, Grushin, Aji, Burkov, GoswamiTewari, Franz, Miransky, Pesin, Landsteiner, Zyuzin} and experimental~\cite{Exp1, Exp2, Exp3, Exp5,Exp8} focus have largely been centered around simple Weyl systems, possessing pairs of (anti-)monopole with unit charge ($n=1$). However, various condensed matter systems endow an unprecedented opportunity to explore the territory of \emph{multi-Weyl semimetals}, characterized by pairs of (anti)-monopole of arbitrary integer charge $n$~\cite{DaiDoubleWeyl, BernevigDoubleWeyl, HasanDoubleWeyl, Nagaosa}. The quasiparticle dispersion for any $n>1$ possesses a natural anisotropy, as displayed in Fig.~\ref{spectra}. But, underlying discrete rotational symmetry in a lattice imposes a strict restriction on the available monopole charge in real materials, namely $|n| \leq 3$~\cite{Nagaosa}. Thus far most of the known examples of Weyl materials have $n=1$~\cite{ArmitageReview, HasanARCMP, FelserARCMP}. Nevertheless, Weyl points with $n=2$ (known as double-Weyl nodes) can in principle be found in HgCr$_2$Se$_4$~\cite{DaiDoubleWeyl, BernevigDoubleWeyl} and SrSi$_2$~\cite{HasanDoubleWeyl}, and A(MoX)$_3$ (with A=Rb, Tl; X=Te) can accommodate Weyl points with $n=3$ (known as triple-Weyl nodes)~\cite{tripleWeyl}. We also note that charge-neutral BdG-Weyl quasiparticles with $n=2$ can also be found in superconducting states of $^{3}$He-A~\cite{VolovikBook}, URu$_2$Si$_2$~\cite{GoswamiBalicas}, UPt$_3$~\cite{GoswamiNevidomskyy}, SrPtAs~\cite{Sigrist}, YPtBi~\cite{royfoster}, for example. Therefore, unveiling the imprint of chiral anomaly in general Weyl semimetals, besides its genuine fundamental importance, is also experimentally pertinent. In this article we study longitudinal magnetotrasport (LMT) in multi-Weyl semimetals, in the \emph{semi-classical} regime. More specifically, resorting to the \emph{kinetic theory} we compute the total out of equilibrium longitudinal magnetoconductivity (LMC) in the parameter regime $T \ll \sqrt{B} \ll \mu$, where $T$ is the temperature and $\mu$ is the chemical potential, measured from the Weyl nodes. 
Note that semiclassical theory of transport is applicable in a parameter regime where quantum corrections can be neglected. In our analysis $T \ll \mu$, and hence the chemical potential or Fermi momentum sets the infrared cutoff in the system. The semiclassical theory is then applicable when $\sqrt{B} \ll \mu$. By contrast, if $T \gg \mu$ then semiclassical appraoch is valid when $\sqrt{B} \ll T$ \cite{PhysRevD.91.125014}.
Manifestation of chiral anomaly in thermal transport for neutral BdG-Weyl quasiparticles is, however, left as a subject for a future investigation.

Kinetic theory can capture the longitudinal magnetotransport in the \emph{weak} magnetic field limit when $\omega_c \tau \ll 1$, where $\omega_c$ is the cyclotron frequency and $\tau$ is the average relaxation time, dominantly arising from elastic scattering due to impurities. In the analysis of longitudinal magnetotransport, which necessarily involves charge pump from the left to the right chiral Weyl point, the relaxation time ($\tau$) is set by \emph{backscattering}. In this regime the Landau levels are not sharply formed (justifying the approach based on kinetic theory) and the path between two successive collision is approximately a \emph{straight line}. Therefore, in the semi-classical (or weak magnetic field) regime, $\tau$ is independent (effectively) of the magnetic field strength and we treat it as a phenomenological input in our analysis from outset. By contrast, in the strong magnetic field limit, the path between two successive collision gets sufficiently \emph{curved}, such that $\tau \equiv \tau(B)$ in addition to ($\omega_c/\mu \ll 1$), and the analysis of magnetotransport demands a quantum mechanical analysis~\cite{ArgyresAdams}. We here focus only on the former situation.

We now provide a brief synopsis of our main findings. We here investigate the LMC in a mutli-Weyl system within the framework of semi-classical theory by considering two possible scenarios, when (a) relaxation of both regular and axial charge is controlled by only one effective time scale in the system (see Sec.~\ref{sec:singlescattering}), and (b) there exists two scattering times in the system (see Sec.~\ref{sec:twoscattering}), arising from the inter-valley ($\tau_{inter}$) and intra-valley ($\tau_{intra}$) processes, for example. While $\tau_{inter}$ is responsible for the relaxation of the axial charge, the intra-valley scattering ensures the isotropy of the distribution function. Irrespective of these details, we show that LMC ($\sigma_{jj}$) always increases as $\sigma_{jj} \sim B^2$ for any value of $n$ as well as for any choice of $j=x,y,z$, which can possibly be observed in experiments. Moreover, $\sigma_{jj}$ scales as $n^3$ with the monopole charge (see Sec.~\ref{Sec:LMTGenW}). However, with a single relaxation time in the system, the contribution of chiral anomaly to LMC cannot be separated from the non-anomalous ones (see Sec.~\ref{sec:LMC_singletime}). Such a separation arises quite naturally in the presence of two scattering times in the medium. In particular, we explicitly demonstrate that when the inter-valley scattering time is sufficiently longer than in intra-valley one (i.e. $\tau_{inter} \gg \tau_{intra}$), the postive LMC is dominated by the anomalous contribution, bearing the signature of the chiral anomaly (see Sec.~\ref{sec:LMC_twotime}).

The rest of the paper is organized as follows. In the next section we introduce the low-energy model for a multi-Weyl semimetal and compute the Berry curvature. In Sec.~\ref{SC:kinetictheory}, we discuss the general formalism of kinetic theory in the context of Weyl semimetals. Sec.~\ref{Sec:LMTGenW} is devoted to the longitudinal magnetotransport in a multi-Weyl metal. The concluding remarks and a discussion on related issues are presented in Sec.~\ref{conclusions}. Additional technical details are relegated to the Appendices.


\section{Berry curvature and topology of a multi-Weyl semimetal}~\label{model}

We begin the discussion by computing the Berry curvature and the associated integer topological invariant of a multi-Weyl semimetal, featuring Weyl nodes with arbitrary integer monopole charge $n$. The low-energy Hamiltonian of a multi-Weyl semimetal is given by~\cite{DaiDoubleWeyl, BernevigDoubleWeyl, Nagaosa, RoyBeraSau, RoyGoswamiJuricic} 
\begin{equation}~\label{eq:hmulti}
H_{n} \left( \mathbf{p} \right) = \alpha_{n} p^n_{\bot} \left[ \cos \left( n \phi_{p} \right) \sigma_{x} +\sin \left( n \phi_{p} \right) \sigma_{y} \right] + v p_z \sigma_{z} \equiv \epsilon_{\mathbf{p}} \: \left( \mathbf{n}_{\mathbf{p}} \cdot \boldsymbol{\sigma} \right),
\end{equation}
where $\phi_{p}= \tan^{-1} \left(p_y/p_x \right)$, $p_{\bot} = \sqrt{p^2_x + p^2_y}$, $\mathbf{n}_{\mathbf{p}}= ( \alpha_{n} p^n_{\bot}\cos \left( n \phi_{p} \right), \alpha_{n} p^n_{\bot}\sin \left( n \phi_{p} \right), v p_z )\epsilon^{-1}_{\mathbf{p}}$, and the set of Pauli matrices $\boldsymbol{\sigma}= \left( \sigma_x,\sigma_y,\sigma_z \right)$ operate on the (pseudo-)spin indices. Momentum ${\bf p}$ is measured from the Weyl node. The energy dispersion in the close proximity to a Weyl node is given by $\pm \epsilon_{\mathbf{p}}$, where $\pm$ respectively corresponds to the conduction and valence bands, and 
\begin{equation}\label{eq:espectrum}
\epsilon_{\mathbf{p}} =  \sqrt{ \alpha^2_{n} p^{2 n}_{\bot} + v^2 p^2_z}. 
\end{equation}
The quasiparticle spectra in a multi-Weyl semimetal along various high symmetry directions are shown in Fig.~\ref{spectra}.
Due to the \emph{doubling theorem} Weyl nodes always appear in pairs~\cite{nielsen}, which we refer here as valley degrees of freedom.

\noindent The components of the Berry curvature close to a Weyl node are defined as
\begin{equation}
\Omega^{\pm}_{\mathbf{p} \phantom{.} a}= \pm \frac{1}{4} \epsilon_{a b c} \mathbf{n}_{\mathbf{p}} \cdot \left( \frac{\partial \mathbf{n}_{\mathbf{p}}}{\partial p_b} \times \frac{\partial \mathbf{n}_{\mathbf{p}}}{\partial p_c} \right) ,
\end{equation}
for the conduction ($\mathbf{\Omega}^{+}_{\mathbf{p}}$) and valence ($\mathbf{\Omega}^{-}_{\mathbf{p} }$) bands. For concreteness, we now focus on the conduction band and for brevity take $\mathbf{\Omega}^{+}_{\mathbf{p} } \to \mathbf{\Omega}_{\mathbf{p}}$.
For a multi-Weyl semimetal we then find
\begin{equation}
\mathbf{\Omega}^{(s)}_{\mathbf{p}} =\frac{s}{2} \frac{n v \alpha_n^2 }{\epsilon_{\mathbf{p}}^{3} }p_\perp^{2(n-1)} \: \left( p_x, p_y, n p_z \right), \label{eq:BCC}
\end{equation}
with $s=\pm$ corresponds to two valleys. 
Notice that upon integrating the Berry curvature over a closed surface $\Sigma$, we find the integer topological invariant of a multi-Weyl semimetal
\begin{equation}
s\,n=\frac{1}{2 \pi} \oint_{\Sigma} \mathbf{\Omega}^{(s)}_{\mathbf{p}} \cdot d\mathbf{S},
\label{eq:topological}
\end{equation}
where $d\mathbf{S}$ is the differential area vector (see Appendix~\ref{ap:topological} for details). Therefore, the integer topological invariant of a Weyl node measures the amount of Berry flux enclosed by a unit area surface, and the Weyl nodes act as source and sink of Abelian Berry curvature of strength $n$.

At this point it is worth pausing to appreciate the dimensionality of various physical quantities in the natural units, in which we set $\hbar=c=k_B=1$. Here the Fermi velocity ($v_{_F}$) plays the role of the velocity of light ($c$). In units of energy, the electric charge has dimension \emph{zero}, while electric and magnetic fields have dimensions \emph{two}, $v$ is \emph{dimensionless} and $\alpha_{n}$ has dimension $1-n$ ~\footnote{While the Fermi velocity $v_F$ and $\alpha_1$ are dimensionless in the natural unit, $\alpha_n$ for $n \geq 2$ bears the dimension of (energy)$^{1-n}$, such that $\alpha_n k^n_\perp$ has the dimension of energy. Note that $\alpha_1$ and $\alpha_2$ are respectively the Fermi velocity and the inverse mass of gapless Weyl excitations in the $xy$-plane. However, there is no standard nomenclature for $\alpha_n$ with $n>2$. }. At last, the central quantity of this study, the conductivity, has dimension \emph{one}, as guaranteed by the gauge invariance.


\section{Kinetic Theory}~\label{SC:kinetictheory}

Kinetic theory is a semiclassical framework, which we employ for the rest of our analysis. We assume the following hierarchy of scales $T \ll \sqrt{B} \ll \mu$, where $T$ is temperature, $B$ is the magnetic field and $\mu$ is the chemical potential, measured from the band-touching point. In this regime, one can ignore the Landau quantization and use Boltzmann kinetic equation
\begin{equation}
	\partial_t f^{(s)}+ \nabla_{\mathbf{x}} f^{(s)} \cdot \dot{\mathbf{x}}^{(s)} + \nabla_{\mathbf{p}} f^{(s)} \cdot \dot{\mathbf{p}}^{(s)} =C[ f^{(s)}],
	\label{eq:B1}
\end{equation}
which describes the evolution of the particle distribution function $f^{(s)}$ in the phase space, where $s$ is the valley index and $C\left[ f^{(s)} \right]$ denotes the collision integral. The effective semiclassical dynamics of Weyl quasiparticles is modified by the Berry curvature in  momentum space, which leads to the following equations of motion
\begin{align}
\dot{\mathbf{x}}^{(s)} {}= \mathbf{v}_\mathbf{p} + \dot{\mathbf{p}}^{(s)} \times \mathbf{\Omega}^{(s)}_{\mathbf{p}}, \quad
\dot{\mathbf{p}} ^{(s)}{}= e\mathbf{E} + e \dot{\mathbf{x}}^{(s)} \times \mathbf{B},
\end{align} 
where $\mathbf{v}_\mathbf{p}=\nabla_{\mathbf{p}} \epsilon_{\mathbf{p}}$ is the group velocity~\cite{Sundaram&Niu}. A comment about the energy dispersion ($\epsilon_{\mathbf{p}}$) is due at this stage. In this article we take $\epsilon_{\mathbf{p}}$ to be the dispersion relation obtained from the effective Hamiltonian [see Eq.~(\ref{eq:espectrum})]. Wave-packet construction reveals that a correction proportional to the inner product of the wave-packet orbital magnetization and the magnetic field should be added to the standard energy dispersion $\epsilon_{\mathbf{p}}$~\cite{RevModPhys.82.1959}. 
In particular for $n=1$ Weyl semi-metals, Lorentz invariance requires $\epsilon_{\mathbf{p}}\to \epsilon_{\mathbf{p}} \mp \frac{1}{2p^2}\mathbf{p}\cdot \mathbf B $~\cite{Son:2012zy,Chen:2014cla}. However, such a correction can lead to an undesired consequence: group velocity becomes bigger than the Fermi velocity $v_f$~\cite{anomalousMaxwell}, as $v_f$ plays the role of the velocity of light in our construction. On the other hand, in~\cite{Zyuzin} the LMC was studied for Weyl semi-metals in the context of kinetic theory concluding that such a modification in the dispersion relation only changes the LMC \emph{quantitatively}, without altering its overall $B^2$ dependence. Therefore, considering the issues with the group velocity and the conclusions of~\cite{Zyuzin}, we neglect this correction in the present article and leave its imprint on LMC as a subject for a future investigation.

 The challenge to solve Eq.~\eqref{eq:B1} arises from the complex form of the collision term, which captures the interactions between particles. Nevertheless, significant progress can be made by employing the so-called \emph{relaxation time approximation}, which encodes the fact that the system returns to equilibrium via scattering events among its constituent particles and impurities \cite{soto2016kinetic}. This process is controlled by a phenomenological parameter which can be interpreted as the average time between two successive collisions. The nature of collisions should follow as a physical input and different choices correspond to different physical outcomes. Specifically, we here analyse two different collision integrals and the corresponding physical scenarios in two subsequent sections. Most importantly we assume that in the semiclassical limit the average scattering times can be considered to be independent of the magnetic field strength for the following reason: in the weak field limit, the radius of the cyclotron orbit is so large that the path between to successive collisions can be approximated as a straight line, and concomitantly $B$-independent. We also assume the relaxation time to be independent of the angles.

 
\subsection{Collisions with single effective relaxation time}~\label{sec:singlescattering}

Our first choice of the collision term assumes the existence of a single relaxation time ($\tau$). The collision integral then takes the following form
\begin{align}\label{Eq:collsinglet}
C_1[ f^{(s)} ]=- \frac{\delta f^{(s)} }{\tau}\,,
\end{align}
where $\delta f^{(s)}=f^{(s)}-f_0$, and $f_0$ the equilibrium Fermi-Dirac distribution function. This type of collision integral was recently used in Refs.~\cite{Gorbar:2017toh,Hidaka:2017auj,Rybalka:2018uzh}.    
The above collision integral has to be taken with care as it assumes that impurity scattering relaxes both regular and axial charge densities \cite{soto2016kinetic}. Therefore we assume the equilibrium state is given by fixed electron and vanishing axial chemical potentials.  Such a scenario is common for open systems, an example given by electronic systems in the presence of charged impurities. However, we here do not derive the above collision integral from any microscopic model, rather treat it as a phenomenological input in the kinetic theory formalism.
To show this explicitly we calculate the semiclassical expressions for the chiral currents (${\mathbf J}^{(s)}$). First we invert the semiclassical equations of motion and obtain~\cite{Duval,Loganayagam:2012pz,Stephanov&Yin}
\begin{align}\label{eq:velocity}
\dot{\mathbf{x}}^{(s)} {}&= \left( 1+e \mathbf{B}\cdot \mathbf{\Omega}^{(s)}_{\mathbf{p}} \right)^{-1} \left[ \mathbf{v}_\mathbf{p} + e \mathbf{E} \times \mathbf{\Omega}^{(s)}_{\mathbf{p}}+ e \left(\mathbf{v}_\mathbf{p} \cdot \mathbf{\Omega}^{(s)}_{\mathbf{p}}\right)\mathbf{B} \right],\\
\dot{\mathbf{p}}^{(s)} {}&= \left( 1+e \mathbf{B}\cdot \mathbf{\Omega}^{(s)}_{\mathbf{p}} \right)^{-1} \left[e\mathbf{E} + e \mathbf{v}_\mathbf{p} \times \mathbf{B}+ e^2 \left(\mathbf{E} \cdot \mathbf{B}\right)\mathbf{\Omega}^{(s)}_{\mathbf{p}} \right].
\label{eq:EOM}
\end{align}
The presence of the Berry curvature modifies the phase space volume element by the factor $ (1+e \mathbf{B}\cdot \mathbf{\Omega}^{(s)}_{\mathbf{p}})$, which satisfies the Liouville equation \cite{Stephanov&Yin}
\begin{equation}
	\partial_t (1+e \mathbf{B}\cdot \mathbf{\Omega}^{(s)}_{\mathbf{p}})+ \nabla_{\mathbf{x}}\cdot \left[(1+e \mathbf{B}\cdot \mathbf{\Omega}^{(s)}_{\mathbf{p}}) \dot{\mathbf{x}}^{(s)}\right] + \nabla_{\mathbf{p}}\cdot \left[(1+e \mathbf{B}\cdot \mathbf{\Omega}^{(s)}_{\mathbf{p}}) \dot{\mathbf{p}}^{(s)}\right] = 2\pi s\, n\,e^2\mathbf E\cdot \mathbf B\,\delta^3(\mathbf p)\,.
\end{equation}
Combining the last expression with the Boltzmann equation [Eq.~(\ref{eq:B1})] we arrive at the following continuity equation
\begin{equation}\label{eq:nonconsereq}
\partial_t\rho^{(s)} + \nabla\cdot\mathbf J^{(s)} = \frac{s\, e^3\,n}{4\pi^2} \,\mathbf E\cdot \mathbf B -\frac{\delta\rho^{(s)}}{\tau}\, ,
\end{equation} 
where the charge ($\rho$) and current ($\mathbf{J}$) densities are respectively defined as
\begin{eqnarray}
\rho^{(s)} = e \int \frac{d^3\mathbf{p}}{(2 \pi)^3} \left( 1+e \mathbf{B}\cdot \mathbf{\Omega}^{(s)}_{\mathbf{p}}\right) f^{(s)}, \quad 
\mathbf{J}^{(s)} = e \int \frac{d^3\mathbf{p}}{(2 \pi)^3} \left( 1+e \mathbf{B}\cdot \mathbf{\Omega}^{(s)}_{\mathbf{p}}\right)\dot{\mathbf{x}} f^{(s)}\,.
\end{eqnarray}
Notice that Eq.~(\ref{eq:nonconsereq}) already discerns the connection between the Berry curvature and chiral anomaly, and will imply relaxation of both electromagnetic and axial charges.

Our goal here is to study the system in a homogeneous and stationary state\footnote{Notice that in order to achieve a steady state,  axial charge needs to be relaxed by the presence of impurities, otherwise the parallel electric and magnetic fields would pump charges indefinitely into the system and the LMC would be infinite. }. Therefore, linearising the Boltzmann equation we obtain  
 \begin{equation}~\label{sol:distrF}
 \delta f^{(s)} = - \frac{\tau}{1+e \mathbf{B}\cdot \mathbf{\Omega}^{(s)}_{\mathbf{p}}} \left[ \left( e\mathbf{E} +e^2 \left( \mathbf{E} \cdot \mathbf{B}\right)\mathbf{\Omega}^{(s)}_{\mathbf{p}} \right) \cdot {\mathbf{v}}_{\mathbf{p}} \right] \frac{\partial  f_0}{\partial \epsilon_{\mathbf{p}}} = -\tau e  \mathbf{E}\cdot \dot{\mathbf{ x}}^{(s)} \, \frac{\partial  f_0}{\partial \epsilon_{\mathbf{p}}},
 \end{equation}
to the leading order. The out-of-equilibrium distribution function is proportional to the work done by the electric field between successive collisions. The injected energy is used by the system in two different mechanisms:

$\bullet$ {\bf Transport of charge}. The first term in Eq.~(\ref{sol:distrF}) is proportional to the work done by the electric field to move the electrons along a trajectory with effective velocity $\mathbf{v_p}$.

$\bullet$ {\bf Creation of charges via the anomaly}. Eq.~(\ref{eq:nonconsereq}) suggests that the second term in Eq.~(\ref{sol:distrF}) is proportional to the induced charge, $\delta \rho \sim n \tau \mathbf{E} \cdot \mathbf{B}$.

Hence we can split the out-of-equilibrium distribution function as $\delta f^{(s)} = \delta f^{(s)}_{\rm O}  + \delta f^{(s)}_{\rm A} $, where
\begin{equation}\label{eq:deltafs}
\delta f^{(s)}_{\rm O} = - \frac{e\tau (\mathbf{E}\cdot \mathbf{v_\mathbf p} )}{1+e \mathbf{B}\cdot \mathbf{\Omega}^{(s)}_{\mathbf{p}}}  \frac{\partial  f_0}{\partial \epsilon_{\mathbf{p}}},\qquad
\delta f^{(s)}_{\rm A} = - \frac{ e^2\tau(\mathbf{\Omega}^{(s)}_{\mathbf{p}}  \cdot {\mathbf{v}}_{\mathbf{p}} )}{1+e \mathbf{B}\cdot \mathbf{\Omega}^{(s)}_{\mathbf{p}}}  \frac{\partial  f_0}{\partial \epsilon_{\mathbf{p}}} \left(  \mathbf{E} \cdot \mathbf{B}\right)\,.
\end{equation}

Now the current operator can be decomposed as $\mathbf{J}^{(s)} =  \mathbf{J_{O}}^{(s)}+\mathbf{J_{AH}}^{(s)}+\mathbf{J_{CM}}^{(s)}$~\cite{Stephanov&Yin},
where $\mathbf{J_{O}}^{(s)}$, $\mathbf{J_{AH}}^{(s)}$ and $\mathbf{J_{CM}}^{(s)}$ denotes the Ohmic, anomalous Hall and out-of-equilibrium chiral magnetic currents, respectively. For simplicity we will assume the multi-Weyl metal is made of two valleys, therefore the specific form of each component reads
\begin{eqnarray}
\mathbf{J_{O}} &=& e \sum_{s=\pm}\int \frac{d^3\mathbf{p}}{(2 \pi)^3} \mathbf{v}_\mathbf{p}  \,\delta f^{(s)}, \\
\mathbf{J_{AH}} &=&e^2\mathbf{E} \times\sum_{s=\pm}\, \int \frac{d^3\mathbf{p}}{(2 \pi)^3}  \mathbf{\Omega}^{(s)}_{\mathbf{p}} \,\delta f^{(s)}, \\
\mathbf{J_{CM}} &=& e^2 \mathbf{B}\sum_{s=\pm}\,\int \frac{d^3\mathbf{p}}{(2 \pi)^3} \left(\mathbf{v}_\mathbf{p} \cdot \mathbf{\Omega}^{(s)}_{\mathbf{p}}\right)\,\delta f^{(s)}\,.
\end{eqnarray}
As we are interested in computing the LMC, and the anomalous Hall current is always transverse to the electric field, we ignore it from now on. Note that each component of the current receives two sub-contributions, which can be appreciated by expressing them as

\begin{eqnarray}\label{eq:cmc}
\mathbf{J_{O}} &=& e\,\sum_{s=\pm}\int \frac{d^3\mathbf{p}}{(2 \pi)^3} \mathbf{v_p}\left[  \frac{4\pi^2e\delta\rho^{(s)}}{sn} \frac{ \left(\mathbf{v}_\mathbf{p} \cdot \mathbf{\Omega}^{(s)}_{\mathbf{p}}\right) }{1+e \mathbf{B}\cdot \mathbf{\Omega}^{(s)}_{\mathbf{p}}} +  \frac{e\tau(\mathbf{E}\cdot \mathbf{v_\mathbf p} ) }{1+e \mathbf{B}\cdot \mathbf{\Omega}^{(s)}_{\mathbf{p}}}  \right] \left(-\frac{\partial  f_0}{\partial \epsilon_{\mathbf{p}}}\right) \,\\
\mathbf{J_{CM}} &=& e^2\,\mathbf{B}\sum_{s=\pm}\int \frac{d^3\mathbf{p}}{(2 \pi)^3} \left[  \frac{4\pi^2e\delta\rho^{(s)}}{sn} \frac{ \left(\mathbf{v}_\mathbf{p} \cdot \mathbf{\Omega}^{(s)}_{\mathbf{p}}\right)^2 }{1+e \mathbf{B}\cdot \mathbf{\Omega}^{(s)}_{\mathbf{p}}} +  \frac{e\tau(\mathbf{E}\cdot \mathbf{v_\mathbf p} ) \left(\mathbf{v}_\mathbf{p} \cdot \mathbf{\Omega}^{(s)}_{\mathbf{p}}\right)}{1+e \mathbf{B}\cdot \mathbf{\Omega}^{(s)}_{\mathbf{p}}}  \right] \left(-\frac{\partial  f_0}{\partial \epsilon_{\mathbf{p}}}\right) \,.\nonumber
\end{eqnarray}

The first term is proportional to an imbalance of charge, which causes a net current along the direction of the group velocity or the external magnetic field. The second contribution is related to the energy needed to transport the particles with an effective velocity given by $\mathbf{v_\mathbf {p}}$. Nonetheless, we would like to emphasize that even though we identify an anomalous contribution in the current  with this collision integral [see Eq.~(\ref{Eq:collsinglet})], it is not possible to disentangle it from the non-anomalous one, due to the presence of a single effective scattering time $\tau$. However, if we introduce two different scattering times in the collision integrals, which can arise from inter- and intra-valley scattering processes, then it is conceivable to isolate the anomalous contributions from the non-anomalous one, specifically when $\tau_{inter} \gg \tau_{intra}$, which we discuss in the next section.


\subsection{Collisions with inter-valley and intra-valley relaxation times}~\label{sec:twoscattering}

\begin{figure}[t!]
	\centering
	\includegraphics[width=0.8\textwidth]{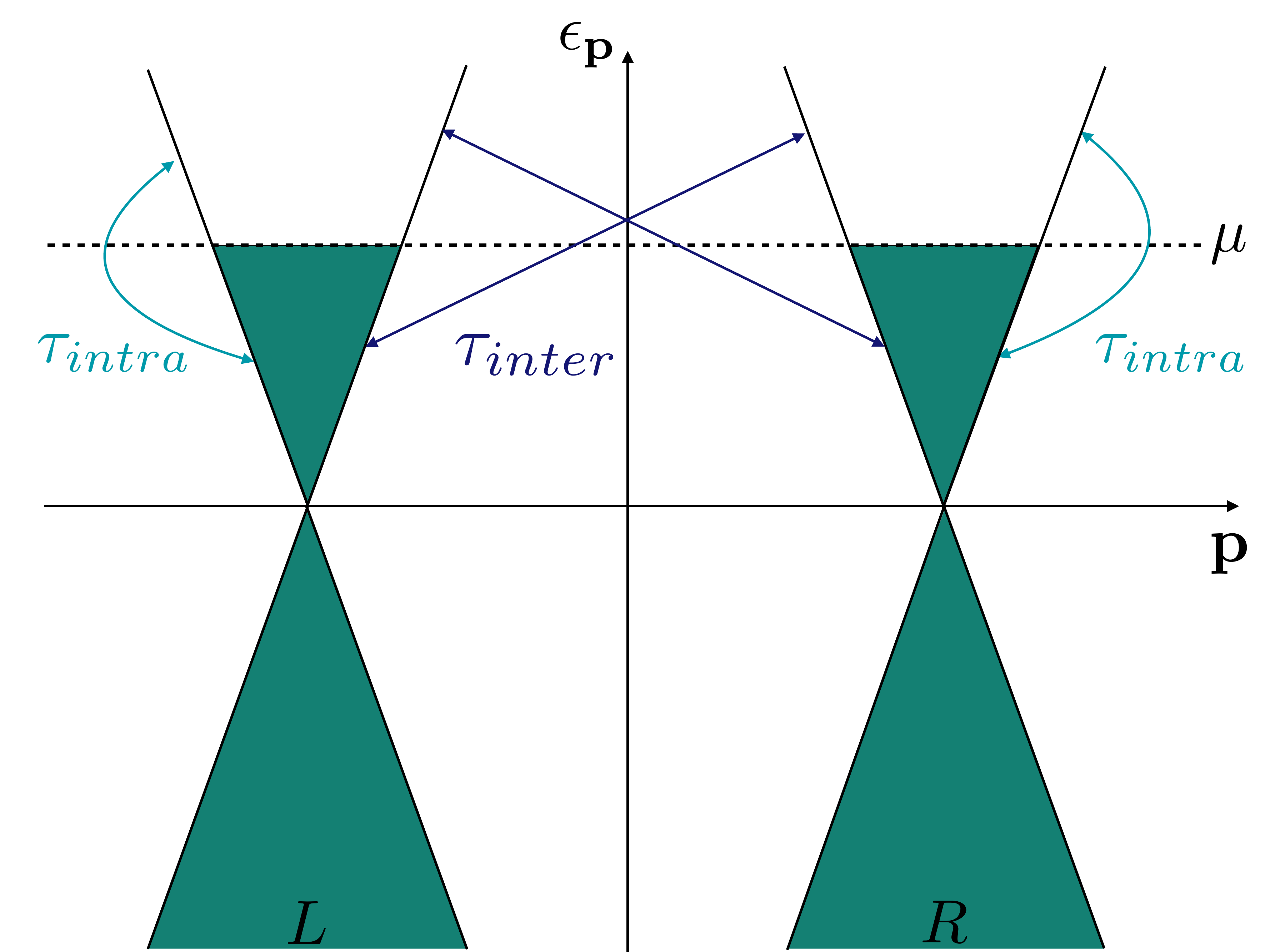}
	\caption{A schematic representation of two scattering processes in a simple Weyl metal, possessing linear dispersion along all three direction. Respectively the forward (intravalley) and back (intervalley) scattering processes are shown by turquoise and blue arrows. The chemical potential ($\mu$) is measured from the band touching point. This construction is also applicable for arbitrary monopole charge $n$. Here $L$ and $R$ respectively denotes the Weyl node with left and right chirality.}
	\label{pic:cones}
\end{figure}

In this subsection we introduce a different collision integral that corresponds to the situation in which there exist two relaxation times. The collision with impurities can either change the chirality of the particle or keep it intact. The former process is captured by the so-called inter-valley relaxation time and the latter one by the intra-valley relaxation time (see Fig.~\ref{pic:cones}).  The inter-valley scattering changes the relative number of particles between two valleys, and is responsible for the ``\emph{charge-pump}" between them. It involves a large momentum transfer and is assumed to be dominated by elastic scattering of particles from impurities. In particular, Gaussian impurities can be a microscopic source of such an inter-valley scattering, while Coulomb impurities, at least in the weak field limit, give rise to forward or intra-valley scattering. Formally we may write the collision term as \cite{Zyuzin}
\begin{equation}
\label{eq:Ake}
C_2[ f^{(s)}]= \frac{\bar{f} ^{(s)} -f^{(s)}}{\tau_{intra}} + \frac{\bar{f} ^{(\bar{s})} -f ^{(s)}}{\tau_{inter}} \equiv \frac{\bar{f} ^{(s)} -f^{(s)}}{\tau^*} + \Lambda^{(s)},
\end{equation}
\noindent where $\tau^{*}=\tau_{inter}\tau_{intra}/\left( \tau_{inter}+\tau_{intra} \right)$, $\bar{s}=-s$, 
\begin{equation}
\bar{f} ^{(s)} = \left\langle \left( 1+e \mathbf{B}\cdot \mathbf{\Omega}^{(s)}_{\mathbf{p}}\right) f^{(s)} \right\rangle, \quad 
\Lambda^{(s)}=\frac{\bar{f} ^{(\bar{s})} -\bar{f} ^{(s)}}{\tau_{inter}}. \nonumber
\end{equation}  
The angular brackets stand for a generalized average over the angles ($\theta$ and $\phi$) 
\begin{equation}
\left\langle \ldots \right\rangle= \frac{\Gamma(\frac{1}{2}+\frac{1}{n})}{2 \pi^{3/2} \, \Gamma(\frac{1}{n})}\int d\phi d\theta \left( \sin \theta \right)^{2/n-1} \ldots, 
\end{equation}
introduced in the new coordinate system 
\begin{equation}
p_x = \left( \epsilon_{\mathbf{p}} \frac{\sin \theta}{\alpha}\right)^{1/n} \cos \phi, \quad p_y = \left( \epsilon_{\mathbf{p}} \frac{\sin \theta}{\alpha}\right)^{1/n} \sin \phi, \quad p_z =  \epsilon_{\mathbf{p}} \frac{\cos \theta}{v},
\end{equation}
compatible with the symmetry of the problem. 

After introducing this average the phase space volume integral reads
\begin{equation}
\int \frac{d^3 p}{(2 \pi)^3}\ldots = \frac{2 \pi^{3/2} \, \Gamma(\frac{1}{n})}{nv\Gamma(\frac{1}{2}+\frac{1}{n})}\int \left(\frac{\epsilon_{\mathbf{p}}}{\alpha_n}\right)^{2/n}\frac{\mathrm{d}\epsilon_{\mathbf{p}}}{(2\pi)^3}\langle\ldots\rangle\,.
\end{equation}
Given this new collision integral the continuity equation can be written as follows
\begin{equation}\label{eq:nonconsereq2}
\partial_t\rho^{(s)} + \nabla\cdot\mathbf J^{(s)} = \frac{s\, e^3\,n}{4\pi^2} \,\mathbf E\cdot \mathbf B -\frac{s}{2}\frac{ \rho_5}{\tau_{inter}}\,.
\end{equation} 
From the above equation, after writing the corresponding electromagnetic and axial continuity equations, it can be seen how $\tau_{inter}$ relaxes only the axial charge $\rho_5=(\rho^{(+)}-\rho^{(-)})/2$.

By solving the kinetic equation we obtain the following leading order solution for the distribution function
\begin{align}
f^{(s)}= \bar{f}^{(s)}+ \tau^{*} \left(\Lambda^{(s)} - \dot{\mathbf{p}}^{(s)} \cdot \mathbf{v} \, \p_{\epsilon_{\mathbf{p}}}f_0\right),
\end{align}
where $\Lambda^{(s)}$ can be obtained by averaging the product of the phase space measure with the kinetic equation (see Appendix~\ref{ap:twotimes} for the detailed computations), leading to 
\begin{equation}
\Lambda^{(s)}= s e^2 \left(\mathbf{E}\cdot\mathbf{B} \right) n^2 v  \left( \frac{\epsilon}{\alpha} \right)^{-2/n} \frac{\Gamma(\frac{1}{2}+\frac{1}{n})}{ \pi^{1/2} \, \Gamma(\frac{1}{n})}  \p_{\epsilon_{\mathbf{p}}}f_0. 
\end{equation}
Finally, the expression for the vector current (considering only a pair of nodes) reads as
\begin{align}
\mathbf{J}{}&= e^2\tau_{inter}\mathbf{B}  \int \frac{d^3 p}{(2 \pi)^3}  \left(\mathbf{v}_{\mathbf{p}}\cdot\mathbf{\Omega}^{(+)}_{\mathbf{p}} \right) \,  \Lambda^{(-)}  -2e^2\tau^*\mathbf{B}  \int \frac{d^3 p}{(2 \pi)^3}  \left(\mathbf{v}_{\mathbf{p}}\cdot\mathbf{\Omega}^{(+)}_{\mathbf{p}} \right)   \Lambda^{(-)}\nonumber\\ 
{}&+ e \tau^{*}  \sum_{s=\pm}\int \frac{d^3 p}{(2 \pi)^3} \left[ \mathbf{v}_{\mathbf{p}}+e \left(\mathbf{v}_{\mathbf{p}}\cdot\mathbf{\Omega}_{\mathbf{p}}^{(s)} \right) \mathbf{B} \right]   \left(- \dot{\mathbf{p}}^{(s)} \cdot \mathbf{v}_{\mathbf{p}} \, \p_{\epsilon_{\mathbf{p}}}f_0\right)\label{eq:current2} .
\end{align}
The first term in the above expression for the current corresponds to the LMC computed in Refs.~\cite{Son&Spivak,Spivak&Andreev,Burkov2} for $n=1$ Weyl semimetals, which becomes the dominant once we take $\tau_{inter} \gg \tau_{intra}$. The rest of the contributions are associated to the effective relaxation time $\tau^*$. Notice that the second line coincides with Eqs.~\eqref{eq:cmc} after setting $\tau \to \tau^*$. Now we proceed to the computation of LMC with the above two collision integrals.


\section{Magnetotransport in the multi-Weyl system}~\label{Sec:LMCGen}~\label{Sec:LMTGenW}

Previous studies reporting a positive LMC in a simple Weyl semimetal (with $n=1$), solely computed the contribution which has a simple connection to the chiral magnetic effect. Here we show that even in the general case with higher monopole charge (with $n>1$), the LMC is possitive for both collision integrals. Otherwise, the LMC ($\sigma_{jj}$) can be computed from the following definition
\begin{equation}
\sigma_{jj}=\sum_{s=\pm }\frac{\partial {\bf J}^{s}}{\partial E_j} \cdot \hat{j},
\end{equation}
where $\hat{j}$ is the unit vector in the $j^{th}$ direction. The electric and magnetic fields are assumed to have the following form $\mathbf E = E\hat j$ and $\mathbf B = B\hat j$. We here present the analysis for two different physical scenarios corresponding to the collision integrals [see Eq.~\eqref{Eq:collsinglet} and \eqref{eq:Ake}].

\subsection{LMC with single effective relaxation time}~\label{sec:LMC_singletime}

A single relaxation time does not distinguish between the processes relaxing the axial and vector currents. As a result LMC receives contributions from both chiral magnetic and Ohmic processes. For convenience we split the total conductivity  as follows
\begin{equation}
{\sigma}_{jj} = 2 \, \sigma^{(1)}_{jj; \tau}+\sigma^{(2)}_{jj; \tau} + \sigma^{(3)}_{jj; \tau},
\end{equation}
where various components ($\sigma^{(k)}_{jj}$) in the above equation are given by the following integral expressions
\allowdisplaybreaks[4]
 \begin{align}
 \sigma ^{(1)}_{jj; \tau} {}&= \tau e^3 B \sum_{s=\pm }\int \frac{d^3\mathbf{p}}{(2 \pi)^3}  \frac{ (\mathbf{v}_\mathbf{p})_j ( \mathbf{\Omega}^{(s)}_{\mathbf{p}} \cdot \mathbf{v}_{\mathbf{p}})}{1+e \mathbf{B}\cdot \mathbf{\Omega}^{(s)}_{\mathbf{p}}} \left(- \frac{\partial  f_0}{\partial \epsilon_{\mathbf{p}}} \right)  \label{eq:s1},\\
\sigma ^{(2)}_{jj; \tau} {}&= \tau e^4 B^2 \sum_{s=\pm }\int \frac{d^3\mathbf{p}}{(2 \pi)^3}  \frac{ ( \mathbf{\Omega}^{(s)}_{\mathbf{p}} \cdot \mathbf{v}_{\mathbf{p}})^2}{1+e \mathbf{B}\cdot \mathbf{\Omega}^{(s)}_{\mathbf{p}}}\left(- \frac{\partial  f_0}{\partial \epsilon_{\mathbf{p}}} \right) \label{eq:s2}, \\
\sigma ^{(3)}_{jj; \tau} {}&= \tau e^2  \sum_{s=\pm }\int \frac{d^3\mathbf{p}}{(2 \pi)^3}  \frac{ (\mathbf{v}_\mathbf{p})^2_j }{1+e \mathbf{B}\cdot \mathbf{\Omega}^{(s)}_{\mathbf{p}}} \left(- \frac{\partial  f_0}{\partial \epsilon_{\mathbf{p}}} \right) \label{eq:s3}.
 \end{align}

\begin{figure}%
 	\centering
 	\subfigure[]{%
 		\label{fig:cond1}%
 		\includegraphics[width=0.50\textwidth]{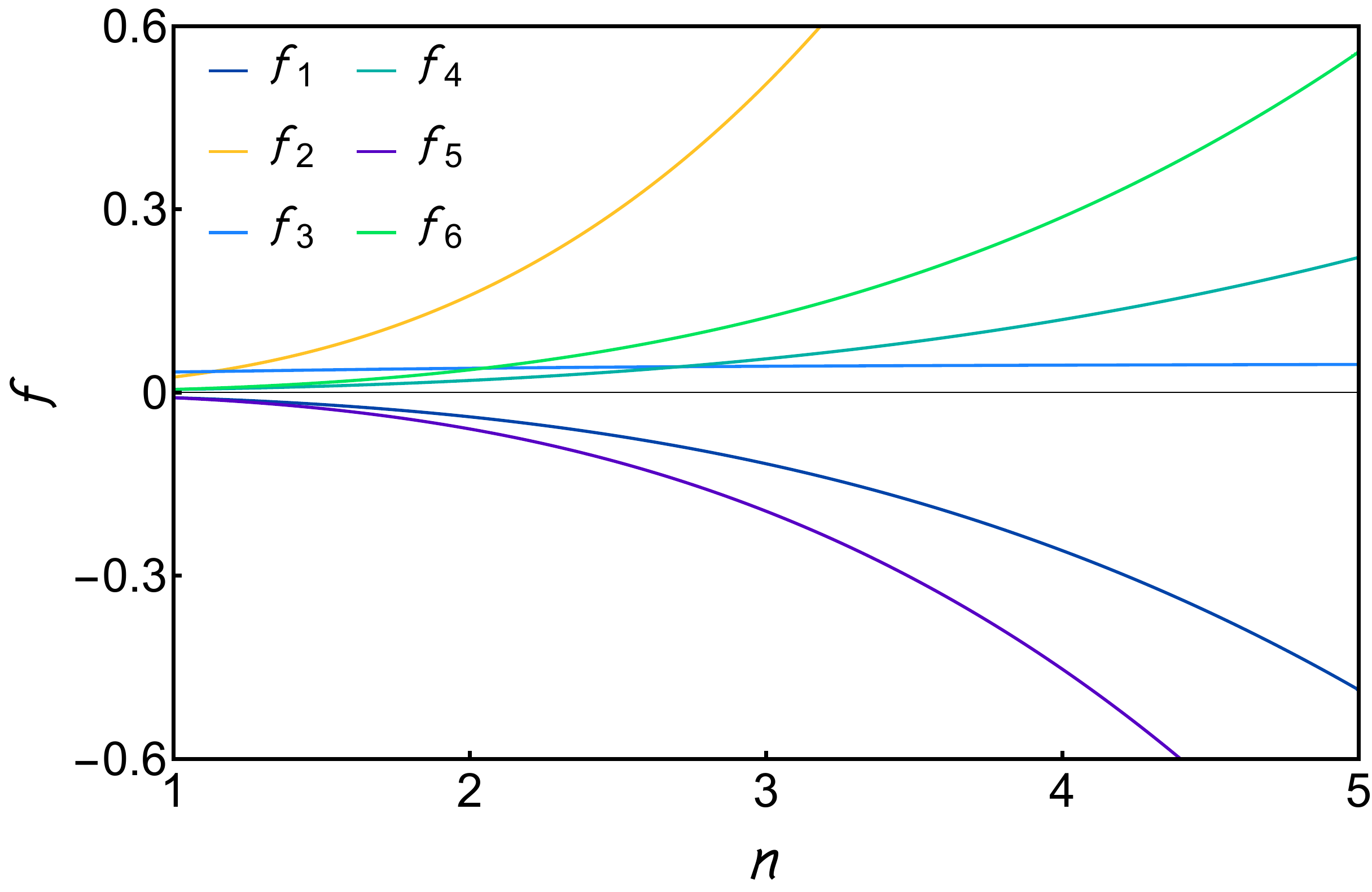}}%
 	\quad
 	\subfigure[]{%
 		\label{fig:cond2}%
 		\includegraphics[width=0.465\textwidth]{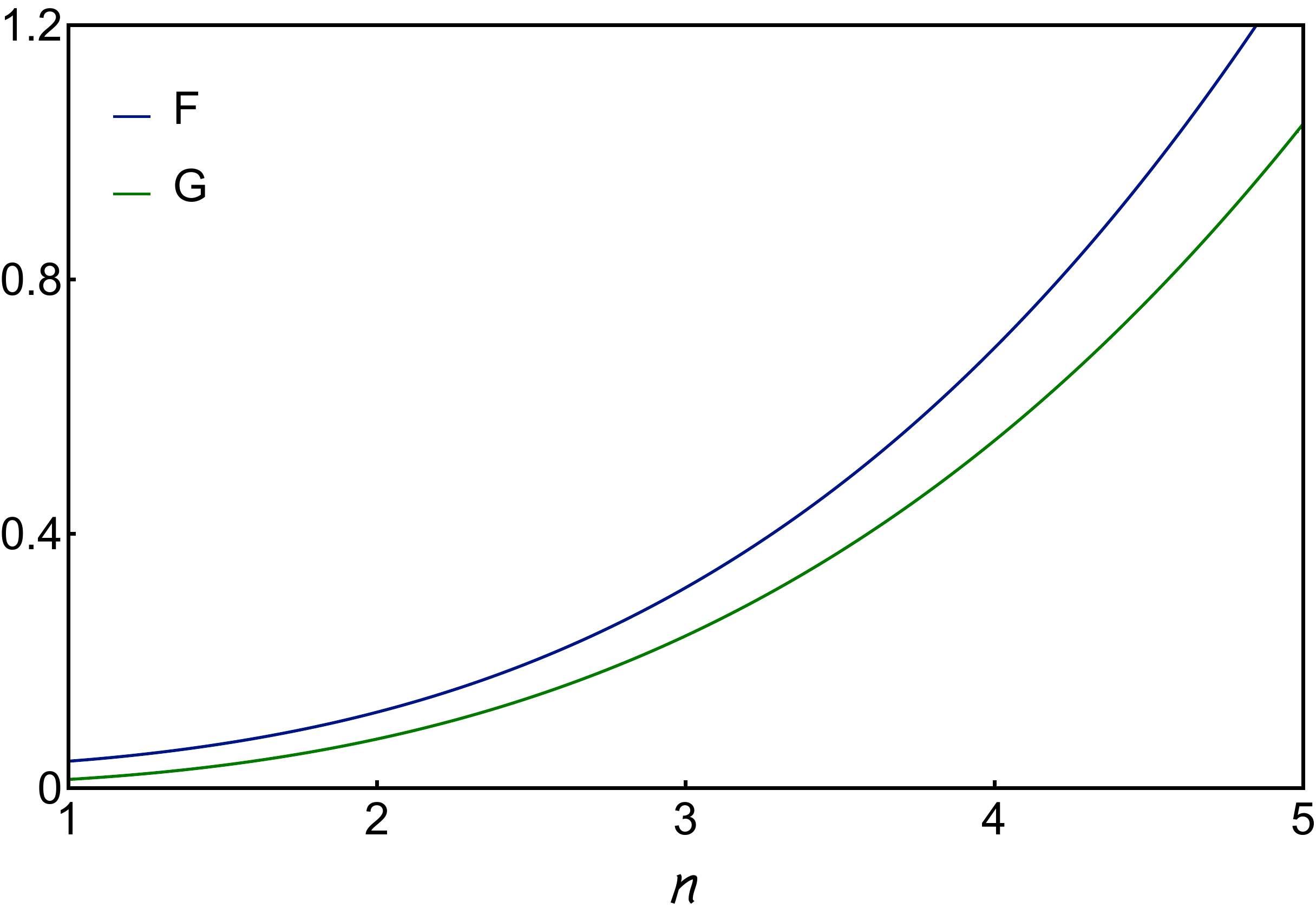}}%
 	\caption{Scaling of the functions (a) $f_{j} (n)$s for $j=1, \cdots, 6$, defined in Eqs.~\eqref{eq:f1234} and \eqref{eq:f56} and (b) $F (n)$ and $G (n)$, respectively appearing in Eqs.~\eqref{eq:sigmazzt} and \eqref{eq:sxxa}, with the monopole's charge $n$, for $n \in \left[ 1,5\right]$.}
 	\label{fig:Conductivities}
 \end{figure}

\noindent Next we compute the components of $\sigma_{jj}$ for various choices of $j$ (for concreteness we choose $j=x,y,z$) for arbitrary $n$ (monopole charge of the Weyl node).   
A generalization of the simple Weyl-node metal involves a momentum space merging of $n$ simple Weyl points with the \emph{same} chirality at a specific point in the momentum space. This situation is qualitatively similar to the ones in two-dimensional bilayer (for $n=2$) and trilayer (for $n=3$) graphene, where respectively the bi-quadratic and bi-cubic touching of the valence and conduction bands can be considered as merging of two and three momentum space vortices. As a result a defect in the form of double- and triple-vortex is realized in these two systems, respectively~\cite{Cstro-RMP}. The low energy dispersion can then be characterized by multi-Weyl nodes with linear dispersion only along one high symmetry direction and $n^{th}$ polynomial dispersion along the remaining two directions. For concreteness, the linear dispersion is chosen to be along the $z$-direction. We seek to understand how such spectral anisotropy manifest in LMC and, in particular, how does it affect the response from anomalies. In what follows we thus compute the LMC along the $z$ direction and perpendicular to it (in the $x-y$ plane). First we consider the situation where $\mathbf{E}=E \hat{z}$ and $\mathbf{B}=B \hat{z}$. Following the steps highlighted above (see Appendices~\ref{ap:anistropic} and~\ref{ap:Convergence} for details) we find the various components of LMC [defined in Eqs.~(\ref {eq:s1})-(\ref{eq:s3})] to be
 \begin{equation}
  \sigma^{(1)}_{zz; \tau}= f_1(n) \; \sigma^{n}_0, \quad 
 \sigma^{(2)}_{zz; \tau} =  f_2(n) \; \sigma^{n}_0, \quad
 \sigma^{(3)}_{zz; \tau} = f_3 (n) \; \sigma^{n}_M + f_4(n) \; \sigma^{n}_0,\label{eq:szza}
 \end{equation} 
 where 
 \begin{eqnarray}~\label{eq:f1234}
 f_1(n) &=& -\frac{n^3 \Gamma \left[ 2-\frac{1}{n}\right]}{16 \pi^{3/2}\Gamma \left[ \frac{7}{2} -\frac{1}{n}\right] }, \quad
 f_2(n)=\frac{n^3 \Gamma \left[ 2-\frac{1}{n}\right]}{8 \pi^{3/2}\Gamma \left[ \frac{5}{2} -\frac{1}{n}\right] }, \quad 
 \nonumber \\
 f_3(n) &=& \frac{\Gamma\left[1+\frac{1}{n} \right]}{4 \pi^{3/2}\Gamma\left[\frac{3}{2}+\frac{1}{n}\right]}, \qquad 
 f_4(n)=\frac{3 n^3 \Gamma\left[2-\frac{1}{n} \right]}{32 \pi^{3/2}\Gamma\left[\frac{9}{2}-\frac{1}{n}\right]},
 \end{eqnarray}
 and 
 \begin{equation}
 \sigma^{n}_0= v \left( \frac{\alpha_n}{\mu}\right)^{2/n} \tau e^4 B^2, \quad 
 \sigma^{n}_M= v \left( \frac{\alpha_n}{\mu} \right)^{-2/n} \tau e^2
 \end{equation}
 bear the dimensionality of conductivity for any value of $n$, and respectively they capture the magnetoductivity and metallic conductivity. The scaling of the functions $f_j(n)$s are shown in Fig.~\ref{fig:cond1}. In the above expression we kept the terms only up to the order $B^2$. Therefore, the total LMC along the $z$ direction in a multi-Weyl system is given by 
 \begin{equation}
 \sigma_{zz}= \left[ 2 f_1(n)+f_2(n)+f_4(n) \right] \sigma^{n}_0 + f_3(n) \sigma^{n}_M \equiv  F(n) \sigma^{n}_0 + f_3(n) \sigma^{n}_M.
\label{eq:sigmazzt} \end{equation}  
 The scaling of the function $F(n)$ is shown in Fig.~\ref{fig:cond2} (blue curve). Finally we align the electric and magnetic fields along the $\hat{x}$ direction. Following the exact same steps we immediately find
 \begin{equation}\label{eq:sigmaxx1234}
  \sigma^{(1)} _{xx;\tau}= f_5(n) \sigma^{n}_0, \quad
 \sigma^{(2)} _{xx;\tau} = \sigma^{(2)} _{zz;\tau}, \quad
 \sigma^{(3)} _{xx;\tau} = \frac{\tau e^2 n \mu^2}{6 \pi^2 v} +f_6(n) \sigma^{n}_0, 
 \end{equation}
 where 
 \begin{equation}\label{eq:f56}
 f_5(n)=-\frac{ n^3 \Gamma \left[ 3- \frac{1}{n}\right]}{16 \pi^{3/2}\Gamma \left[ \frac{7}{2}- \frac{1}{n}\right]}, \quad 
 f_6(n)=\frac{3 n^3 \Gamma\left[ 4-\frac{1}{n}\right]}{64 \pi^{3/2}\Gamma\left[ \frac{9}{2}-\frac{1}{n}\right]}.
 \end{equation}
 Scaling of $f_5 (n)$ and $f_6 (n)$ with $n$ is shown in Fig.~\ref{fig:cond1} . Hence, the total LMC along the $x$ direction in a multi-Weyl system is given by 
 \begin{equation}
 \sigma_{xx}= \left[ 2 f_5(n)+f_2(n)+f_6(n) \right] \sigma^{n}_0 + \frac{\tau e^2 n \mu^2}{6 \pi^2 v} \equiv  G(n) \sigma^{n}_0 +  \frac{\tau e^2 n \mu^2}{6 \pi^2 v}. \label{eq:sxxa}
 \end{equation}  
 The scaling of the function $G(n)$ is shown in Fig.~\ref{fig:cond2} (green curve). Due to an in-plane rotational symmetry, $\sigma^{(i)}_{xx}=\sigma^{(i)}_{yy}$, implying $\sigma_{xx}=\sigma_{yy}$.

 We now discuss the results. Notice that the contribution to the LMC solely arising from  $\sigma^{(2)}_{jj;\tau}$, is independent of the choice of $j=x,y,z$. Such a behaviour unveils the underling topological protection of the chiral anomaly on LMC.  However, the rest of the contributions to the LMC, namely $\sigma^{(1)}_{jj;\tau}$ and $\sigma^{(3)}_{jj;\tau}$  scales differently along various high symmetry directions. This should be contrasted with the linear $n$ dependence of the equilibrium chiral magnetic conductivity. As a result the total LMC $\sigma_{jj}$ despite being always positive, is direction dependent [compare Eq.~\eqref{eq:sigmazzt} and Eq.~\eqref{eq:sxxa}]. We also note that for weak enough magnetic field the leading contribution to LMC goes as $B^2$, irrespective of the direction. Otherwise, $\sigma_{jj}$ scales as $n^3$ with the monopole charge of the Weyl nodes.

We would like to make a final remark regarding the positive LMC. This observable is believed to be a direct indication of the  underling chiral anomaly. However, to the best of our knowledge there is no solid proof of that statement (directly connecting negative LMR arising from the Berry curvature with the quantum chiral anomaly computed from the triangle diagrams)\footnote{In Ref.~\cite{PhysRevLett.119.166601} a positive LMC is discused in a context without Weyl nodes.}~\cite{PhysRevD.73.025017, PhysRevD.97.051901, PhysRevD.97.016018}. Nonetheless, upon splitting the LMC of the multi-Weyl semimetal in terms of the out-of-equilibrium chiral magnetic and  the Ohmic\footnote{For comparative reasons we ignore the Drude part in the Ohmic conductivity.} conductivities (see Eq. \eqref{eq:cmc})
 \begin{eqnarray}
 \sigma^{CM}_{zz} &=& 
  \frac{n^3 \Gamma \left[3-\frac{1}{n} \right]}{16 \pi^{3/2}\Gamma \left[\frac{7}{2}-\frac{1}{n} \right]}\sigma_0^n, \qquad\qquad \sigma^{O}_{zz} =  
 -\frac{n^3 \Gamma \left(3-\frac{1}{n}\right)}{32 \pi ^{3/2} \Gamma \left(\frac{9}{2}-\frac{1}{n}\right)}\sigma_0^n\,,\\
 \sigma^{CM}_{xx} &=& 
  \frac{n^2 \left( 3n-1\right) \Gamma \left[2-\frac{1}{n} \right]}{32 \pi^{3/2}\Gamma \left[\frac{7}{2}-\frac{1}{n} \right]}\sigma_0^n,\qquad   \sigma^{O}_{xx} = 
 -\frac{(5 n-1) n^2 \Gamma \left(3-\frac{1}{n}\right)}{128 \pi ^{3/2} \Gamma \left(\frac{9}{2}-\frac{1}{n}\right)}\sigma_0^n\,,
 \end{eqnarray}
we observe that the dominant LMC for arbitrary $n$ is the one related to the chiral magnetic conductivity. This supports the idea of a direct relation between positive LMC and the chiral anomaly. in the next section we show that in presence of two distinct time scales it possible to demonstrate a one-to-one correspondence between LMC and the chiral anomaly when $\tau_{inter} \gg \tau_{intra}$.

 \subsection{LMC with two relaxation times}~\label{sec:LMC_twotime}
 
We now present the expression for magnetotransport when both intervalley and intravalley scattering times are taken into account. This corresponds to the collision term, shown in Eq.~\eqref{eq:Ake}. In this case the computation reduces to the evaluation of only the first line of Eq. \eqref{eq:current2} [see Appendix~\ref{ap:twotimes} for details], since the second line identically matches with the expression for LMC for the single relaxation time collision integral after changing $\tau\to\tau^*$. Therefore the LMC for the actual case can be written down as follows
\begin{align}~\label{eq:LMC_twotimes}
\sigma_{jj} &= \tau_{inter}  \frac{e^4  n^3 v \Gamma(\frac{1}{2}+\frac{1}{n})}{4\pi^{5/2}\Gamma(\frac{1}{n})}\left( \frac{\alpha}{\mu }\right)^{2/n} B^2 \\
{}& - \tau^*\frac{e^4  n^3 v \Gamma(\frac{1}{2}+\frac{1}{n})}{2\pi^{5/2}\Gamma(\frac{1}{n})}\left( \frac{\alpha}{\mu }\right)^{2/n} B^2 +  2 \, \sigma^{(1)}_{jj;\tau^*}+\sigma^{(2)}_{jj;\tau^*}+\sigma^{(3)}_{jj;\tau^*}  \nonumber,
\end{align}
 where $ \{\sigma^{(1)}_{\tau^{*}},\sigma^{(2)}_{\tau^{*}},\sigma^{(3)}_{\tau^{*}}\} $ are given by Eqs.~\eqref{eq:szza}-\eqref{eq:f56}. In this case we see that when $\tau^{*} \ll \tau_{inter}$, corresponding to $\tau_{inter} \gg \tau_{intra}$, we obtain the generalization of the LMC of Ref.~\cite{Son&Spivak} for the multi-Weyl case. In this limit, the LMC is purely governed by the chiral anomaly, which is direction independent and thus topological in nature.



\section{Discussion and Conclusions}~\label{conclusions}

To summarize, we here present a comprehensive analysis of LMC in a three-dimensional multi-Weyl semimetal in the semi-classical regime, which can be accessed in experiments for sufficiently weak magnetic field, such that $\omega_c \tau \ll 1$ (thus no Landau quantization). The distribution of the underlying Berry curvature in the momentum space is isotropic only when $n=1$, for which the dispersion of Weyl fermions scales linearly with all three components of momentum. By contrast, due to a natural anisotropy in the Weyl dispersion for $n >1$ [see Fig.~\ref{spectra}], the system looses Lorentz invariance and the Berry curvature is no longer uniformly distributed [see Sec.~\ref{model}]. In this work we investigate the imprint of the (anisotropic) Berry curvature on LMC in multi-Weyl system.

Throughout we assume the electric and magnetic fields to be parallel to each other. Specifically, we considered two different types of collision integrals corresponding to two different physical scenarios: (a) When both regular and axial charge are relaxed by a \emph{single} effective scattering time ($\tau$) [see Sec.~\ref{sec:singlescattering}], and (b) in the presence of both inter-valley and intra-valley scattering processes, respectively characterized by $\tau_{inter}$ and $\tau_{intra}$ [see Sec.~\ref{sec:twoscattering} and Fig.~\ref{pic:cones}]. In the latter construction only $\tau_{inter}$ causes the relaxation of the axial charge.

Within the framework of single scattering time approximation, we show that the contribution to LMC arising from the chiral anomaly gets mixed with the non-anomalous ones, and they cannot be separated [see Sec.~\ref{sec:LMC_singletime}]. By contrast, these two contributions are separated when we invoke two different time-scales in the collision integrals in the form of inter-valley ($\tau_{inter}$) and intra-valley ($\tau_{intra}$) scattering times [see Sec.~\ref{sec:LMC_twotime}]. In particular, when $\tau_{inter} \gg \tau_{intra}$ the dominant contribution to LMC arises from chiral anomaly [see Eq.~\eqref{eq:LMC_twotimes}] and it is proportional to the inter-valley scattering time $\tau_{inter}$. However, irrespective of these details we show that the LMC always increases as $\sigma_{jj} \sim B^2$ for $j=x,y,z$, and scales as $n^3$ with the monopole charge. While in the single scattering time approximation the amplitude of $\sigma_{jj}$ is always direction dependent, LMC becomes direction independent in the presence of two scattering times, but only when $\tau_{inter} \gg \tau_{intra}$. In this regime LMC solely arises from the chiral anomaly, and its direction independence reveals its topological origin. In brief, our work strongly suggests an one-to-one correspondence among the underlying Berry curvature of the Weyl medium, the chiral anomaly and the positive LMC in a multi-Weyl system. The proposed topologically robust LMC can be observed in Weyl systems at weak magnetic fields, if back-scattering dominates over the forward one (yielding $\tau_{inter}\gg\tau_{intra}$), which can be realized when concentration of Gaussian impurities is sufficiently larger than that for Coulomb impurities. We also note that for sufficiently weak magnetic field the \emph{weak anti-localization} effect leads to a negative LMC~\cite{antilocalization}. The interplay of chiral anomaly and weak anti-localization effects and the crossover behaviour between them remains an unresolved issue at this moment.

Finally, we wish to draw a comparison between our conclusions regarding the LMC in a multi-Weyl system in the weak field and the one obtained in a quantum limit ($\omega_c \tau \gg 1$)~\cite{xiaolibitan}, when Landau levels are sharply formed (strong magnetic field regime). In the strong field limit it has been demonstrated that positive LMC scales \emph{linearly} with the monopole charge ($n$) and magnetic field ($B$), as long as it is applied along the $\hat{z}$ direction (separating two Weyl nodes). The linear-dependence of positive LMC on $n$ comes from the fact that the zeroth Landau level in a multi-Weyl semimetal possesses an exact and topologically protected $n$-fold degeneracy~\cite{roy-sau}. In a simple Weyl semimetal ($n=1$) such a linear dependence on the B-field is insensitive to its direction. However, for $n=2$ and $3$ as one tilts the field away from the $\hat{z}$ direction the  LMC (still positive) starts to develop a \emph{non-linear} dependence on the $B$-field, and most likely scales as $B^2$ when the field is aligned in the $x-y$ plane. Such a stark distinct crossover behaviour of LMT from semi-classical to quantum regime (accessed by systematically increasing the strength of the magnetic field or strength of impurity scattering) along various direction of a multi-Weyl system is extremely fascinating, which can also be observed in real materials by tilting the magnetic field away from high symmetry directions.

\acknowledgments

B. R. is thankful to Nordita for hospitality. P. S. is supported by the Deutsche Forschungsgemeinschaft via the Leibniz Programme. We thank Dam Thanh Son for discussions.


\appendix

\section{Computation of Berry Curvature}~\label{ap:topological}

In this appendix we elaborate on the computation of the integer topological invariant of generalized Weyl semimetals. To proceed with the analysis we exploit the azimuthal symmetry of the system for $n>1$. First, we express the Berry curvature in cylindrical coordinates according to
\begin{align}
\mathbf{\Omega}_{\mathbf{p}} = \frac{n \alpha_n^2 v p^{2n-1}_{\bot}}{2 (\alpha_n^2 p^{2n}_{\bot}+v^2 p^2_z)^{3/2}} (\mathbf{e}_{p_\bot}+n p_z p^{-1}_{\bot}\mathbf{e}_{p_z}).
\end{align}

\begin{figure}[t!]
	\centering
	\includegraphics[width=0.5\textwidth]{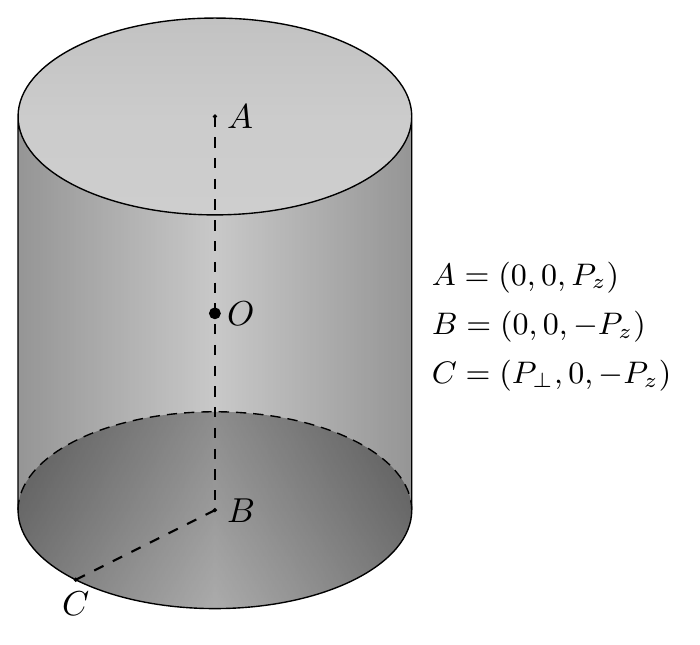}
	\caption{Illustration of the chosen surface for the computation of flux of the Berry curvature in a multi-Weyl semimetal (see Appendix~\ref{ap:topological}). The Weyl monopole is placed at $O$.}
	\label{fig:Cylinder}
\end{figure}

\noindent Then, choosing the surface $\Sigma$ to be a cylinder centred around the monopole (see Fig.~\ref{fig:Cylinder}), we obtain
\\

\begin{align}
\oint_{\Sigma} \mathbf{\Omega}_{\mathbf{p}} \cdot d\mathbf{S} {}&= \int_{\Sigma_S} \mathbf{\Omega}_{\mathbf{p}} \cdot d\mathbf{S_S} +\int_{\Sigma_T} \mathbf{\Omega}_{\mathbf{p}} \cdot d\mathbf{S_T} +\int_{\Sigma_B} \mathbf{\Omega}_{\mathbf{p}} \cdot d\mathbf{S_B} \nonumber\\
{}&= \int_{-P_z}^{P_z} \frac{ \pi n \alpha_n^2 v P^{2n}_{\bot}}{\left[\alpha^2_n P^{2n}_{\bot}+v^2 p^2_z\right]^{3/2}}dp_z + 2 \int_{0}^{P_{\bot}} \frac{ \pi n^2 \alpha_n^2 v p^{2n-1}_{\bot} P_z}{\left[\alpha^2_n p^{2n}+v^2 P^2_z\right]^{3/2}}dp_{\bot} \:\:\:
= 2 \pi n .
\end{align}

\noindent Note that $\Sigma_S$, $\Sigma_T$ and $\Sigma_B$ respectively represents the side ($S$), top ($T$) and bottom ($B$) surfaces of the cylinder.


\section{Calculation of magnetoconductance with two relaxation times}~\label{ap:twotimes}

In this Appendix, we display details of the computation of LMC in the presence of two scattering time in the collision integral [see Eq.~\eqref{eq:Ake}]. We begin with the kinetic equation
\begin{align}
\label{eq:AAke}
\p_t f^{(s)} + \dot{\mathbf{x}}^{(s)} \cdot \p_{\mathbf{x}}f^{(s)}+ \dot{\mathbf{p}}^{(s)} \cdot \p_{\mathbf{p}}f^{(s)} {}&= \frac{\bar{f} ^{(s)} -f^{(s)}}{\tau^*} + \frac{\bar{f} ^{(\bar{s})} -\bar{f} ^{(s)}}{\tau_{inter}} \\
{}&\equiv \frac{\bar{f} ^{(s)} -f^{(s)}}{\tau^*} + \Lambda^{(s)},
\end{align}
\noindent where $\bar{f} ^{(s)} = \left\langle \left( 1+ e \mathbf{B}\cdot\mathbf{\Omega}^{(s)}\right) f^{(s)} \right\rangle$. The angular brackets stand for a generalized average over the angles, to be specified below.

Given the symmetry of the system, we work with the coordinate system in which the radial component corresponds to the energy dispersion relation
\begin{equation}
\epsilon_{\mathbf{p}}= \sqrt{\alpha^2 (p_x^2+p_y^2)^n + v^2 p_z^2},
\end{equation}
\noindent defined by
\begin{align}
\label{eq:coordsyst}
p_x {}= \left( \epsilon_{\mathbf{p}} \frac{\sin \theta}{\alpha}\right)^{1/n} \cos \phi,
p_y {}= \left( \epsilon_{\mathbf{p}} \frac{\sin \theta}{\alpha}\right)^{1/n} \sin \phi, 
p_z {}=  \epsilon_{\mathbf{p}} \frac{\cos \theta}{v} .
\end{align}

\noindent In this coordinate system, the group velocity and the Berry curvature take the simple form
\begin{align}
\mathbf{v}_{\mathbf{p}}{}= \frac{1}{h_1} \hat{\mathbf{\epsilon}}_{\mathbf{p}}, \quad 
\mathbf{\Omega}_{\mathbf{p}} {}= \frac{n^2 v \alpha^2 }{2 \epsilon_{\mathbf{p}}^2 } \left( \frac{\epsilon_{\mathbf{p}} \sin \theta}{\alpha} \right)^{2(n-1)/n} h_1 \hat{\mathbf{\epsilon}}_{\mathbf{p}},
\end{align}
where $h_1= \sqrt{\frac{\cos^2 \theta}{v^2} + \frac{1}{n^2 \epsilon^2_{\mathbf{p}}} \left(\frac{\epsilon_{\mathbf{p}} \sin \theta }{ \alpha}\right)^{2/n}}$.  Finally, the above mentioned average of a quantity $g$ over the angles is defined as follows

\begin{equation}
\left\langle g \right\rangle= \frac{\Gamma(\frac{1}{2}+\frac{1}{n})}{2 \pi^{3/2} \, \Gamma(\frac{1}{n})}\int d\phi d\theta \left( \sin \theta \right)^{2/n-1} g.
\end{equation}

To compute the explicit form of $ \Lambda^{(s)}$ for static and homogeneous solutions we angle-average the kinetic equation after multiplying it by the phase space measure, yielding  

\begin{align}
\label{eq:Ake2}
\left\langle \left( 1+ e \mathbf{B}\cdot\mathbf{\Omega}^{(s)}_{\mathbf{p}}\right) \dot{\mathbf{p}}^{(s)} \cdot \p_{\mathbf{p}}f^{(s)}\right\rangle  {}&= \left\langle \left( 1+ e \mathbf{B}\cdot\mathbf{\Omega}^{(s)}_{\mathbf{p}}\right) \left(  \frac{\bar{f} ^{(s)} -f^{(s)}}{\tau^*} + \Lambda^{(s)} \right) \right\rangle.
\end{align}
\noindent Since $\left\langle \left( 1+ e \mathbf{B}\cdot\mathbf{\Omega}^{(s)}_{\mathbf{p}}\right) \right\rangle=1$, we find
\begin{align}
\label{eq:Ake5}
\Lambda^{(s)}= \left\langle \left( 1+ e \mathbf{B}\cdot\mathbf{\Omega}^{(s)}_{\mathbf{p}}\right) \dot{\mathbf{p}}^{(s)} \cdot \p_{\mathbf{p}}f^{(s)}\right\rangle. 
\end{align}
 Using the equations of motion, we obtain the following simplified expression for $\Lambda^{(s)}$ within the linear response
\begin{align}
\Lambda^{(s)}{}&= \left\langle \left( e \mathbf{E} \cdot \mathbf{v} + e^2 \left(\mathbf{E}\cdot\mathbf{B} \right) \mathbf{\Omega}^{(s)}_{\mathbf{p}}\cdot \mathbf{v}_{\mathbf{p}}  \right) \right\rangle \p_{\epsilon_{\mathbf{p}}}f_0 \nonumber \\
{}&= s e^2 \left(\mathbf{E}\cdot\mathbf{B} \right) n^2 v  \left( \frac{\epsilon_{\mathbf{p}}}{\alpha} \right)^{-2/n} \frac{\Gamma(\frac{1}{2}+\frac{1}{n})}{ \pi^{1/2} \, \Gamma(\frac{1}{n})}  \p_{\epsilon_{\mathbf{p}}}f_0,
\end{align}
\noindent where $f_{0}=\left[ 1+ e^{\beta (\epsilon_{\mathbf{p}}-\mu)}\right]^{-1}$, and $\mu$ is the equilibrium chemical potential. 
The solution of the kinetic equation (in linear response) is given by
\begin{align}
f^{(s)}= \bar{f}^{(s)}+ \tau^{*} \left(\Lambda^{(s)} - \dot{\mathbf{p}}^{(s)} \cdot \mathbf{v}_{\mathbf{p}} \, \p_{\epsilon_{\mathbf{p}}}f_0\right).
\end{align}
The current is defined as \footnote{Note that we here ignore the term responsible for the Hall current.}
\begin{align}
\mathbf{J}^{(s)}{}&= e \int \frac{d^3 p}{(2 \pi)^3}  \left( 1+ e \mathbf{B}\cdot\mathbf{\Omega}^{(s)}_{\mathbf{p}}\right) \dot{\mathbf{x}}^{(s)}f^{(s)}\\
{}&=e \int \frac{d^3 p}{(2 \pi)^3} \left[ \mathbf{v}_{\mathbf{p}}+e \left(\mathbf{v}_{\mathbf{p}}\cdot\mathbf{\Omega}^{(s)}_{\mathbf{p}}\right) \mathbf{B} \right]  \bar{f}^{(s)} \nonumber\\ 
{}&+e \tau^{*} \int \frac{d^3 p}{(2 \pi)^3} \left[ \mathbf{v}_{\mathbf{p}}+e \left(\mathbf{v}_{\mathbf{p}}\cdot\mathbf{\Omega}^{(s)}_{\mathbf{p}} \right) \mathbf{B} \right]   \left(\Lambda^{(s)} - \dot{\mathbf{p}}^{(s)} \cdot \mathbf{v}_{\mathbf{p}} \, \p_{\epsilon_{\mathbf{p}}}f_0\right) \nonumber\\
{}&=e \int \frac{d^3 p}{(2 \pi)^3} \left[ e \left(\mathbf{v}_{\mathbf{p}}\cdot\mathbf{\Omega}^{(s)}_{\mathbf{p}} \right) \mathbf{B} \right] \left( \bar{f}^{(s)} +\tau^* \Lambda^{(s)}\right)\nonumber\\ 
{}&+e \tau^{*} \int \frac{d^3 p}{(2 \pi)^3} \left[ \mathbf{v}_{\mathbf{p}}+e \left(\mathbf{v}_{\mathbf{p}}\cdot\mathbf{\Omega}^{(s)}_{\mathbf{p}} \right) \mathbf{B} \right]   \left(- \dot{\mathbf{p}}^{(s)} \cdot \mathbf{v}_{\mathbf{p}} \, \p_{\epsilon_{\mathbf{p}}}f_0\right).
\end{align}
\noindent For a pair of nodes the vector current can be computed yielding 
\begin{align}
\mathbf{J}{}&= e \sum_{s=\pm } \int \frac{d^3 p}{(2 \pi)^3}  \left( 1+ e \mathbf{B}\cdot\mathbf{\Omega}^{(s)}_{\mathbf{p}}\right) \dot{\mathbf{x}}^{(s)}f^{(s)} \nonumber \\
{}&=e \int \frac{d^3 p}{(2 \pi)^3} \,e \left(\mathbf{v}_{\mathbf{p}}\cdot\mathbf{\Omega}_{\mathbf{p}} \right) \mathbf{B} \left[ s \bar{f}^{(s)} +\bar{s} \bar{f}^{(\bar{s})} +\tau^* \left( s \Lambda^{(s)} + \bar{s}\Lambda^{(\bar{s})}\right)\right] \nonumber\\ 
{}&+ e \tau^{*}  \sum_{s=\pm}\int \frac{d^3 p}{(2 \pi)^3} \left[ \mathbf{v}_{\mathbf{p}}+ e \left(\mathbf{v}_{\mathbf{p}}\cdot\mathbf{\Omega}^{(s)}_{\mathbf{p}} \right) \mathbf{B} \right]   \left(- \dot{\mathbf{p}}^{(s)} \cdot \mathbf{v}_{\mathbf{p}} \, \p_{\epsilon_{\mathbf{p}}}f_0\right) \nonumber\\
{}&=e \int \frac{d^3 p}{(2 \pi)^3} \, e \left(\mathbf{v}_{\mathbf{p}}\cdot\mathbf{\Omega}_{\mathbf{p}} \right) \mathbf{B} \left[ \tau_{inter} \Lambda^{(-)} -2 \tau^* \Lambda^{(-)}\right] \nonumber\\ 
{}&+ e \tau^{*}  \sum_{s=\pm}\int \frac{d^3 p}{(2 \pi)^3} \left[ \mathbf{v}_{\mathbf{p}}+ e \left(\mathbf{v}_{\mathbf{p}}\cdot\mathbf{\Omega}^{(s)}_{\mathbf{p}} \right) \mathbf{B} \right]   \left(- \dot{\mathbf{p}}^{(s)} \cdot \mathbf{v}_{\mathbf{p}} \, \p_{\epsilon_{\mathbf{p}}}f_0\right).
\end{align}
Note that
\begin{align}
{}& e \tau_{inter} \int \frac{d^3 p}{(2 \pi)^3} \left[ e \left(\mathbf{v}_{\mathbf{p}}\cdot\mathbf{\Omega}_{\mathbf{p}} \right) \mathbf{B} \right] \Lambda^{(-)}=\frac{e^4 \tau_{inter} n^3 v \Gamma(\frac{1}{2}+\frac{1}{n})}{4\pi^{5/2}\Gamma(\frac{1}{n})}\left( \frac{\alpha}{\mu}\right)^{2/n} \left( \mathbf{E}\cdot\mathbf{B}\right)\mathbf{B},\\
{}& e \tau^{*} \int \frac{d^3 p}{(2 \pi)^3} \left[ \mathbf{v}_{\mathbf{p}}+e \left(\mathbf{v}_{\mathbf{p}}\cdot\mathbf{\Omega}^{(s)}_{\mathbf{p}} \right) \mathbf{B} \right]   \left(- \dot{\mathbf{p}}^{(s)} \cdot \mathbf{v}_{\mathbf{p}} \, \p_{\epsilon_{\mathbf{p}}}f_0\right)= \mathbf{J}_b,
\end{align}
\noindent where $\mathbf{J}_b$ has the same structure than the vector current computed for $C_1 \left[ f^{(s)}\right]$ but it is now proportional to $\tau^*$. Therefore, the total LMC (in the presence of two scattering times) for a pair of Weyl nodes is given by Eq.~(\ref{eq:LMC_twotimes}).

\section{Computation of magnetoconductivity}~\label{ap:conductivities}

We now present some essential details of the computation of LMC for multi-Weyl semimetal (with $n>1$) that appear in both single and two relaxation time approximations, namely $\sigma^{(k)}_{jj,\tau}$ for $k=1,2,3$ and $j=x,y,z$. Finally, we also justify the power series expansion in powers of $B$ for the calculation of the LMC. 

\subsection{Multi-Weyl semimetal}~\label{ap:anistropic}

The aim of this section is to illustrate how we obtain the results quoted in Eqs.~\eqref{eq:szza}-\eqref{eq:sxxa}. We present the full computation for one of the terms, as the remaining ones can be evaluated in a similar way. Let us focus on $\sigma^{(1)}_{zz,\tau}$ which can be written as
\begin{align}
\sigma ^{(1)}_{zz,\tau} {}&= \frac{\tau e^3 B}{(2 \pi)^2} \sum_{s=\pm}s\int^{\infty}_{0}  d p_{\bot} \int_{-\infty}^{\infty} dp_z \frac{(n^2 \alpha_{n}^2 v^3) p_{\bot}^{2n+1} p_z \:\: \delta (\mu - \sqrt{\alpha_{n}^2 p_{\bot}^{2n}+v^2p^2_z})}{2 p_{\bot}^2 (\alpha_{n}^2 p_{\bot}^{2n} +v^2 p_z^2)^{3/2}+vn^2 \alpha_{n}^2 (s e B)p_{\bot}^{2n} p_z} . \nonumber
\end{align}

\noindent Now we perform the variable substitution $p_z \rightarrow p_z/v $ and $p_{\bot} \rightarrow p_{\bot}   \alpha_{n}^{-1/n} $, yielding
\begin{align}
\sigma ^{(1)}_{zz,\tau} {}&= \frac{\tau e^3 B}{(2 \pi)^2} \sum_{s=\pm}s\int^{\infty}_{0}  d p_{\bot} \int_{-\infty}^{\infty} dp_z \frac{(n^2 \alpha_{n}^{-2/n} v) p_{\bot}^{2n+1} p_z \:\: \delta (\mu - \sqrt{ p_{\bot}^{2n}+p^2_z})}{ 2 \alpha_{n}^{-2/n} p_{\bot}^2 (p_{\bot}^{2n} +p_z^2)^{3/2}+n^2 (se B)p_{\bot}^{2n} p_z}. \nonumber
\end{align}

\noindent Next, we take $p_{\bot} = \tilde{p}_{\bot} ^{1/n}$. For brevity we drop the tildes and find
\begin{align}
\sigma ^{(1)}_{zz,\tau} {}&= \frac{\tau e^3 B}{(2 \pi)^2} \sum_{s=\pm}s\int^{\infty}_{0} d p_{\bot} \int_{-\infty}^{\infty} dp_z \frac{(n \alpha_{n}^{-2/n} v) p_{\bot} p_z \:\: \delta (\mu - \sqrt{ p_{\bot}^{2}+p^2_z})}{ 2 \alpha_{n}^{-2/n} (p_{\bot}^{2} +p_z^2)^{3/2}+n^2 (se B)p_{\bot}^{2(n-1)/n} p_z}. \nonumber 
\end{align}
At last, performing the transformation $p_{\bot}=R \sin{\theta}$ and $p_z=R \cos{\theta}$, we obtain
\begin{align}
\sigma ^{(1)}_{zz,\tau} {}&= \frac{\tau e^3 B (n \alpha_{n}^{-2/n} v)}{(2 \pi)^2}\sum_{s=\pm}s \int^{\pi}_{0} d \theta\int_{0}^{\infty} dR \frac{ R^3 \sin{\theta} \cos{\theta} \:\: \delta (\mu - R)}{2 \alpha_{n}^{-2/n} R^3+ (se B)n^2 (R \cos{\theta})(R\sin{\theta})^{2(n-1)/n}} \nonumber\\
{}&=  \frac{\tau e^3 B (n v)}{2(2 \pi)^2}\sum_{s=\pm} s\int^{\pi}_{0} d \theta \frac{ \sin{\theta} \cos{\theta}}{ 1+ (\frac{se B}{2 \mu^2})n^2 \alpha_{n}^{2/n} \cos{\theta}(\mu \sin{\theta})^{2(n-1)/n}} \label{eq:aint1}.
\end{align}

\noindent The previous coordinates transformations amount to going from the Cartesian coordinates to the ones introduced in Eq.~(\ref{eq:coordsyst}). Performing the same steps, it is straight forward to show that 
\begin{align}
\sigma ^{(2)}_{zz,\tau} {}&=  \frac{ \tau e^4 B^2 (n^3 \alpha_{n}^{2/n} v)}{4(2 \pi)^2 \mu^2}\sum_{s=\pm} \int_{0}^{\pi} d\theta \frac{ \sin \theta  (\mu \sin \theta)^{2(n-1)/n}}{1 +  \left(\frac{se B}{2\mu^2}\right) n^2 \alpha_{n}^{2/n}  \cos \theta  (\mu \sin \theta)^{2(n-1)/n}},\nonumber\\
\sigma^{(3)}_{zz,\tau}{}&=  \frac{\tau e^2 v \mu}{(2 \pi)^2 n \alpha_{n}^{2/n}}\sum_{s=\pm} \int^{\pi}_{0} d \theta \frac{ \cos^2{\theta} (\mu \sin{\theta})^{2/n-1}}{ 1+ \left(\frac{se B}{2 \mu^2}\right)n^2 \alpha_{n}^{2/n} \cos{\theta}(\mu \sin{\theta})^{2(n-1)/n}},\nonumber \\
\sigma^{(1)} _{xx,\tau} {}&=\frac{\tau e^3 B (n^2 \alpha_{n}^{1/n})}{2 (2 \pi)^3 \mu^2} \sum_{s=\pm}s\int_{0}^{2\pi} d\phi \int_{0}^{\pi} d\theta \frac{(\mu \sin\theta)^{3-1/n} \cos \phi}{1+\left( \frac{se B}{2 \mu^3}\right) n v \alpha_{n}^{1/n} \left( \mu \sin \theta\right)^{2-1/n} \cos{\phi} },\nonumber \\
\sigma^{(2)}_{xx,\tau}{}&=\frac{ \tau e^4 B^2 (n^3 \alpha_{n}^{2/n} v )}{4(2 \pi)^3 \mu^3}\sum_{s=\pm} \int_{0}^{2\pi} d\phi \int_{0}^{\pi} d\theta \frac{(\mu \sin \theta)^{3-2/n}}{1+\left( \frac{se B}{2 \mu^3}\right) n v \alpha_{n}^{1/n} \left( \mu \sin \theta \right)^{2-1/n} \cos{\phi} },\nonumber \\
\sigma^{(3)}_{xx,\tau}{}&=\frac{ \tau e^2 n\mu^2}{(2 \pi)^3 v}\sum_{s=\pm} \int_{0}^{2\pi} d\phi \int_{0}^{\pi} d\theta \frac{ \sin^3 \theta \cos^2 \phi}{1+\left( \frac{se B}{2 \mu^3}\right) n v \alpha_{n}^{1/n} \left( \mu \sin \theta \right)^{2-1/n} \cos{\phi} }. \label{eq:aints}
\end{align}

\noindent To compute these integrals, next we need to perform a series expansion of the integrands in powers of $e B/2\mu^2$ (see Appendix \ref{ap:Convergence}). 


\subsection{Power Series Expansion }~\label{ap:Convergence}

\begin{figure}[h]
	\centering 
	\includegraphics[width=0.5\textwidth]{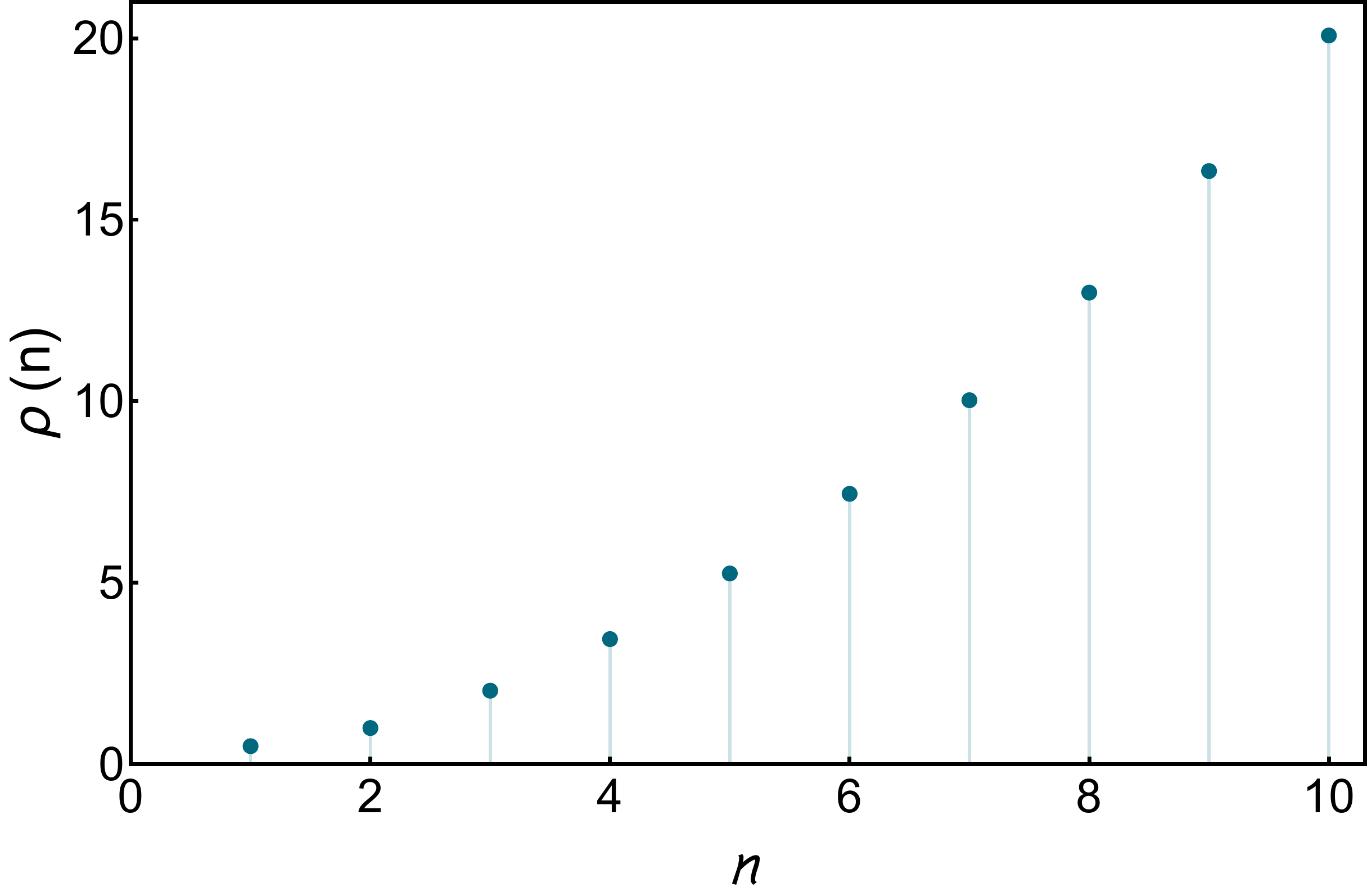}
	\caption{Scaling of $\rho(n)$ [defined in Eq.~\eqref{eq:rho}] with the monopole number $n$ of generalized Weyl fermions.}
	\label{fig:Convergence}
\end{figure}

We seek to perform the integrals from Eqs.~\eqref{eq:aint1} and ~\eqref{eq:aints}. As before, let us focus on  $\sigma^{(1)}_{zz,\tau}$. To compute $\sigma^{(1)}_{zz,\tau}$, we make use of the power series expansion 

\begin{align}
\frac{1}{1+\epsilon}=\sum^{\infty}_{i=0} (-1)^i \epsilon^i,
\end{align}

\noindent  which allows us to write 

\begin{align}
\sigma^{(1)}_{zz,\tau}{}&=  \frac{\tau e^3 B (n v)}{2(2 \pi)^2} \sum_{s=\pm}s\int^{\pi}_{0} d \theta \sum_{i=0}^{\infty}  \sin{\theta} \cos{\theta} (-1)^i \left[ \left( \frac{se B}{2 \mu^2} \right)n^2 \alpha_{n}^{2/n} \cos{\theta}(\mu \sin{\theta})^{2(n-1)/n}\right]^i. \nonumber \\
\end{align}

\noindent To proceed further we need to interchange the integral with the sum sign. A sufficient condition is $\sum_n \int dx |f_n(x)| < \infty $, or equivalently, $\int dx \sum_n |f_n(x)| < \infty $. Let us prove the former condition. First of all, note that

\begin{align}
{}& \int^{\pi}_{0} d \theta  \left| (-1)^i \left[ \left( \frac{e B}{2} \right)n^2 \left( \frac{\alpha_{n}}{\mu}\right)^{2/n} \cos{\theta}( \sin{\theta})^{2(n-1)/n}\right]^i  \sin{\theta} \cos{\theta} \right| \nonumber\\
{}&= \frac{ (e B)^i n^{2i} \left( \frac{\alpha_{n}}{n}\right)^{2i/n}  \Gamma\left[1+\frac{i}{2}\right]\Gamma\left[1+i-\frac{i}{n}\right]}{2^i \Gamma\left[2+i\left( \frac{3}{2} - \frac{1}{n}\right)\right]}.
\end{align}
\noindent Next we compute the ratio 

\begin{align}
r=\lim_{i \rightarrow \infty} \left| \frac{a_{i+1}}{a_{i}} \right|{}& = \lim_{i \rightarrow \infty} \left| \frac{e B}{2 \mu^2}\right| \left| \frac{\alpha_{n}^{2/n}}{\mu^{2/n-2}}\right| \left| \frac{n^{2}  \Gamma\left[\frac{3+i}{2}\right]\Gamma\left[2+i-\frac{1+i}{n}\right] \Gamma\left[2+i\left( \frac{3}{2} - \frac{1}{n}\right)\right]}{2 \Gamma\left[1+\frac{i}{2}\right]\Gamma\left[1+i-\frac{i}{n}\right] \Gamma\left[2+(1+i) \left( \frac{3}{2} - \frac{1}{n}\right)\right]} \right| \nonumber\\
{}&= \left| \frac{e B}{2 \mu^2}\right| \left| \frac{\alpha_{n}^{2/n}}{\mu^{2/n-2}}\right| \lim_{i \rightarrow \infty} \rho_i(n).
\end{align}

\noindent Specifically, for integer $n$ bigger or equal to 1, we have 

\begin{equation}
\rho(n)=\lim_{i \rightarrow \infty} \rho_i(n)=n (n-1)^{(n-1)/n} \left( \frac{3n}{2} -1\right)^{1/n} \left( \frac{n}{3n-2} \right)^{3/2}.
\label{eq:rho}
\end{equation}

\noindent We can now study $\rho (n)$  as a function of $n$ (see Fig.~\ref{fig:Convergence}). For the series to converge $r<1$. With no loss of generality, we can assume $\alpha_{n} /\mu$ to be a positive finite number. Thus, for finite $n$, we can always find the regime where $\left| \frac{e B}{2 \mu^2}\right|< \left(\left| \frac{\alpha_{n}^{2/n}}{\mu^{2/n-2}}\right| \rho(n)\right)^{-1} $, making the series convergent.

It can be shown for all the other terms that we can choose $e B/2 \mu^2$ to be small in order for the series to be convergent. Therefore, we can compute the desired integrals by expanding the integrands in powers of $e B/2 \mu^2$. Keeping only the terms up to quadratic order in the magnetic field, we arrive at the results quoted in Eqs.~\eqref{eq:szza}-\eqref{eq:sxxa}.

\bibliographystyle{JHEP}

\bibliography{MultiNodeWeyl}

\end{document}